\documentclass[12pt]{iopart}
\usepackage{graphics,graphicx,subfigure}

\begin{document}

\title[Effective interaction between guest charges in 2D jellium]{Effective
  interaction between guest charges immersed in 2D jellium}

\author{Ladislav \v{S}amaj}

\address{Institute of Physics, Slovak Academy of Sciences, 
D\'ubravsk\'a cesta 9, SK-84511 Bratislava, Slovakia}
\ead{Ladislav.Samaj@savba.sk}
\vspace{10pt}
\begin{indented}
\item[]
\end{indented}

\begin{abstract}
The model under study is an infinite 2D jellium of pointlike
particles with elementary charge $e$, interacting via the logarithmic
potential and in thermal equilibrium at the inverse temperature $\beta$.
Two cases of the coupling constant $\Gamma\equiv \beta e^2$ are considered:
the Debye-H\"uckel limit $\Gamma\to 0$ and the free-fermion point
$\Gamma=2$.
In the most general formulation, two guest particles, the one with charge
$q e$ (the valence $q$ being an arbitrary integer) and the hard core of
radius $\sigma>0$ and the pointlike one with elementary charge $e$,
are immersed in the bulk of the jellium at distance $d\ge \sigma$.
Two problems are of interest: the asymptotic large-distance behavior of
the excess charge density induced in the jellium and
the effective interaction between the guest particles.
Technically, the induced charge density and the effective interaction
are expressed in terms of multi-particle correlations of the
pure (translationally invariant) jellium system.
It is shown that the separation form of the induced charge density onto
its radial and angle parts, observed previously in the limit $\Gamma\to 0$,
is not reproduced at the coupling $\Gamma=2$.
Based on an exact expression for the effective interaction between
guest particles at $\Gamma=2$, oppositely ($q=0,-1,-2,\ldots$)
charged guest particles always attract one another while likely
($q=1,2,\ldots$) charged guest particles repeal one another up to
a certain distance $d$ between them and then the mutual attraction
takes place up to asymptotically large (finite) distances.
\end{abstract}


\vspace{2pc}


\submitto{\JPA}

\maketitle

\renewcommand{\theequation}{1.\arabic{equation}}
\setcounter{equation}{0}

\section{Introduction} \label{Sec1}
Thermal equilibrium of classical systems of charged particles is of interest
in many branches of physics, especially in soft and condensed matter
\cite{Levin02}.

In the $\nu$-dimensional Euclidean space, the Coulomb potential $v$ at point
${\bf r}=(x_1,x_2,\ldots,x_{\nu})$, induced by a unit charge at the origin
${\bf 0}=(0,0,\ldots,0)$, is defined as the solution of
the $\nu$-dimensional Poisson equation (taken in Gauss units)
\begin{equation}
\Delta v({\bf r}) = - s_{\nu} \delta({\bf r}) ,  
\end{equation}
where $s_{\nu}=2\pi^{\nu/2}/\Gamma(\nu/2)$
($\Gamma$ stands for the Gamma function)
is the surface of the unit sphere in space of dimension $\nu$;
$s_2=2\pi$, $s_3=4\pi$, etc.
The Fourier component of such defined potential behaves like $1/k^2$ which
maintains many generic properties of three-dimensional (3D) Coulomb fluids
with the potential
\begin{equation}
v({\bf r}) = \frac{1}{r} , \qquad r=\vert {\bf r}\vert ,
\end{equation}  
namely perfect screening and the associated sum rules \cite{Martin88}.
In two dimensions (2D), the electric potential, satisfying the boundary
condition $\nabla v({\bf r})\to 0$ as $r\to\infty$, is logarithmic
\begin{equation}
v({\bf r}) = - \ln \left( \frac{r}{r_0} \right) ,
\end{equation}  
where $r_0$ is a free length scale which will be set,
for simplicity, to unity.
2D pointlike charges can be substituted in the 3D space by parallel charged
lines, mimicking 3D polyelectrolytes \cite{Shklovskii99a}.

The weak-coupling (high-temperature) regime of Coulomb systems is described
adequately by the mean-field Poisson-Boltzmann theory \cite{Andelman06},
improved systematically via a loop expansion
\cite{Attard88,Netz00,Podgornik90}, or its linearised Debye-H\"uckel version.
For like-charged spherical colloids, immersed in an electrolyte,
the corresponding DLVO (Derjaguin, Landau, Verwey, Overbeek) theory
\cite{Verwey48} always implies an effective repulsion for
an arbitrary distance between them.
An extension of the 3D DLVO theory to anisotropic colloids
implies an effective interaction potential which remains anisotropic at all
distances between the colloids \cite{Chapot04,Chapot05};
at asymptotically large distances from the colloids, the effective interaction
potential factorizes itself onto its distance-dependent and
angle-dependent parts. 

In the opposite strong-coupling (low-temperature) limit, a counter-intuitive
effective attraction of like-charged colloids was observed at small enough
temperatures, experimentally
\cite{Khan85,Kjellander88,Bloomfield91,Kekicheff93,Dubois98}
as well as by computer simulations
\cite{Gulbrand84,Kjellander84,Gronbech97}; see also reviews
\cite{Boroudjerdi05,Naji13}.
The theoretical treatment of the strong-coupling (SC) regime is not
understood completely yet.
A functional approach leading to a virial fugacity expansion
\cite{Moreira00,Moreira01,Netz01} implies, in the leading SC order,
a ``single-particle picture'' of independent particles moving in
the electric field of the charged walls.
Although the leading SC picture is in agreement with Monte Carlo (MC)
simulations \cite{Moreira00,Moreira01,Moreira02,Kanduc07}, higher correction
orders of the virial SC approach fail to reproduce correctly MC data.
Another approach to the SC regime follows from the existence of classical
Wigner crystals at zero temperature \cite{Shklovskii99b,Levin99,Grosberg02}.
The consideration of harmonic deviations of charged particles from their Wigner
positions \cite{Samaj11a,Samaj11b} leads, in the leading order, correctly
to the single-particle theory.
Moreover, the first correction to the single-particle theory agrees very well
with MC data, in the large range of Coulomb couplings.
There are other Wigner-structure theories \cite{Samaj16,Palaia18}, adapting
the idea of a correlation hole \cite{Nordholm84,Forsman04}, which provide
very good results even for intermediate and small couplings.

In this paper, we consider the 2D one-component Coulomb plasma (jellium)
of mobile pointlike particles of the same (say elementary) charge $e$
which interact pairwisely via the 2D logarithmic potential.
Like in standard jellium systems \cite{Levin02,Baus80}, the neutralizing
uniform background charge density is spread uniformly over the domain
the charged particles are confined to. 
The jellium is in thermal equilibrium at the inverse temperature
$\beta=1/(k_{\rm B}T)$, the only relevant parameter is the coupling
constant $\Gamma=\beta e^2/\varepsilon$ with $\varepsilon$ being
the dielectric constant (set to unity) of the medium.
2D Coulomb systems confined to domains with specific geometries (infinite,
disc, semi-infinite, etc.) are of special importance because they are exactly
solvable, besides the limit $\Gamma\to 0$, also at the free-fermion
coupling $\Gamma=2$ \cite{Jancovici81,Alastuey81},
see reviews \cite{Jancovici92,Forrester98}.
For a {\em finite} (not too large) number of particles $N$,
the partition function and particle densities are expressible explicitly
for the sequence of the coupling constants $\Gamma = 2\gamma$ where
$\gamma=2,\ldots$ is a positive integer.
This can be done by writting the integer powers of Vandermonde determinant
in terms of Jack polynomials \cite{Tellez99,Tellez12} or anticommuting-field
theory formulated on a chain of $N$ sites \cite{Samaj95,Samaj04,Samaj15};
for a relation between the two methods, see Ref. \cite{Grimaldo15}.

The effective interaction between like-charged ions immersed in the 2D
jellium at the free-fermion coupling $\Gamma=2$, in the bulk
as well as near the jellium edge, was studied in Ref. \cite{Ma01}. 
The effective interaction of two pointlike unit ions $e$ can be obtained
explicitly and it is always repulsive.
The situation becomes more complicated when one attaches a hard core of radius
$\sigma$ (i.e., an empty void from which plasma ions are excluded)
to the interacting ions. 
Because a fermion ``overlap'' matrix becomes non-diagonal in the presence of
the guest charges, the corresponding partition function (from which
the effective interaction between the two ions is calculated)
was calculated perturbatively, in the leading order of the expansion in
the hard-core parameter $\sigma$.
It was shown that the effective interaction between guest charges results from
a competition between ion-ion repulsion, which dominates at short distances,
and a longer ranged ion-void attraction which dominates at large distances.
This means that at the coupling $\Gamma=2$ the void of colloids causes
an effective attraction of like charged ions at sufficiently large distances
between colloids.
It is not clear whether this scenario takes place also in higher orders of
the hard-core parameter $\sigma$.
Note that the described mechanism of attraction of like-charged colloids
does not exclude another scenario for larger values of the couplings $\Gamma>2$
where an effective attraction might take place even for pointlike
like-charged ions, at least in some intervals of the distance between the ions.

The most general formulation studied in this paper consists in two guest
particles at distance $d$ immersed in the bulk of the 2D jellium, the one
with charge $q e$ (the valence $q$ being an arbitrary integer) and
the hard core of radius $\sigma>0$ and the pointlike one with elementary
charge $e$.
Although this model is slightly different from the one treated in \cite{Ma01}
(with $q=1$ and a hard core attached to both guest charges), it involves all
relevant physical phenomena of the original problem and its exact solvability 
is crucial.
Two problems are of interest: the asymptotic large-distance behavior of
the charge density induced in the jellium and the effective interaction
between the guest particles.
Technically, the induced charge density and the effective interaction
are expressed in terms of multi-particle correlations of the
pure (translationally invariant) jellium system which are exactly known
at the coupling $\Gamma=2$; this avoids manipulations with the non-diagonal
fermion overlap matrix.
It is shown that the separation form of the induced charge density at
asymptotically large distances from the guest particles onto its radial
and angles parts, observed previously in the Debye-H\"uckel limit $\Gamma\to 0$
\cite{Chapot04,Chapot05}, is not reproduced at the coupling $\Gamma=2$.
Based on an exact expression for the effective interaction between
the guest particles at $\Gamma=2$, oppositely ($q=0,-1,-2,\ldots$)
charged guest particles always attract one another.
On the other hand, likely ($q=1,2,\ldots$) charged guest particles repeal
one another up to a certain distance between them and then
the mutual attraction takes place up to an infinite distance.

The paper is organized as follows.
In section \ref{Sec2}, the 3D Debye-H\"uckel analysis of the induced
charge density \cite{Chapot04,Chapot05} is extended to the guest charges
$q_1 e$ and $q_2 e$ at distance $d$, immersed in the bulk of the 2D jellium.
The separation form of the charge density onto its radial and angle parts
at asymptotically large distances from the guest particles is reproduced.
Section \ref{Sec3} deals with the general 2D formalism at arbitrary
coupling $\Gamma$ for an infinite homogeneous jellium (section \ref{Sec31})
and the jellium perturbed by one (section \ref{Sec32}) or two
(section \ref{Sec33}) guest charges.
The emphasis is put on the particle densities in the presence of one
or two guest charges which are expressed in terms of multi-particle
correlations of the pure (translationally invariant) jellium system.
Section \ref{Sec4} brings a short recapitulation on how to calculate
multi-particle correlation functions for an infinite jellium system
with a circular symmetry of the neutralising background at the coupling
$\Gamma=2$.
The simplest case of two identical guest particles with unit charge
is described in section \ref{Sec5}.
Section \ref{Sec6} deals with the configuration of two distinct guest
charges, the one with charge $q e$ ($q$ being an integer) and the hard core
of radius $\sigma$ and the pointlike one with the elementary charge $e$.
Section \ref{Sec7} is a brief summary with concluding remarks.

\renewcommand{\theequation}{2.\arabic{equation}}
\setcounter{equation}{0}

\section{Linear Debye-H\"uckel theory} \label{Sec2}
Let us put a guest pointlike charge $q e$ (integer $q$ is the valence) say
at the origin ${\bf 0}=(0,0,\ldots,0)$ of an infinite $\nu$-dimensional
jellium system of particles with elementary charge $e$.
Due to the presence of the guest charge, the constant particle density $n_0$
in the interior of the jellium changes itself to a density profile
$n({\bf r}\vert \{ qe,{\bf 0}\})$, the corresponding excess charge density
(of mobile particles minus the background) is given by
\begin{equation}
\rho({\bf r}\vert \{ qe,{\bf 0}\}) = e
\left[ n({\bf r}\vert \{ qe,{\bf 0}\}) - n_0 \right] .
\end{equation}
In the mean-field approach valid at high temperatures,
the particle density $n$ is determined by the Boltzmann factor of
the mean local electric potential $\phi$ as follows
\begin{equation}
n({\bf r}\vert \{ qe,{\bf 0}\}) =
n_0 {\rm e}^{-\beta e\phi({\bf r}\vert \{ qe,{\bf 0}\})} \approx
n_0\left[ 1 - \beta e \phi({\bf r}\vert \{ qe,{\bf 0}\}) + \cdots \right] ,    
\end{equation}
so that $\rho({\bf r}\vert \{ qe,{\bf 0}\})
\approx - \beta e^2 n_0 \phi({\bf r})$.
Inserting this charge density into the Poisson equation
$\Delta\phi=-s_{\nu}\rho({\bf r}\vert \{ qe,{\bf 0}\})$, one gets
\begin{equation} \label{crucial}
\frac{{\rm d}^2\phi}{{\rm d}r^2} + \frac{(\nu-1)}{r}
\frac{{\rm d}\phi}{{\rm d}r} - \kappa^2 \phi = 0 ,
\end{equation}  
where $\kappa=\sqrt{s_{\nu}\beta e^2 n_0}$ is the inverse Debye-H\"{u}ckel
length.
We are looking for the solution which behaves like the pure Coulomb
potential at small distances from the guest charge,
\begin{equation} \label{crucial1}
\phi(r\vert \{ qe,{\bf 0}\}) \mathop{\sim}_{r\to 0} q e v(r) ,
\end{equation}
and vanishes at large distances from the guest charge,
\begin{equation}
\phi(r\vert \{ qe,{\bf 0}\}) \mathop{\sim}_{r\to \infty} 0 .
\end{equation}

Let us now consider two guest particles in the bulk of the jellium system,
at distance $d$ say along the $x$-axis, the one with charge $q_1 e$ at point
${\bf r}_1=(-d/2,0,\ldots,0)$ and the other with charge $q_2 e$ at point
${\bf r}_2=(d/2,0,\ldots,0)$; this configuration mimics an anisotropic colloid.
Since the differential equation for the electric potential (\ref{crucial})
is linear, the superposition of individual solutions for each guest particle
is also its solution, i.e.,
\begin{equation} \label{superposition}
\phi({\bf r}\vert \{ q_1 e,{\bf r}_1\},\{ q_2 e,{\bf r}_2\}) =
\phi({\bf r}-{\bf r}_1\vert \{ q_1 e,{\bf 0}\})
+ \phi({\bf r}-{\bf r}_2\vert \{ q_2 e,{\bf 0}\}) ,
\end{equation}
where the effective potential $\phi$ is the solution of Eq. (\ref{crucial}). 
The counterparts of the asymptotic relation close to the guest charges
(\ref{crucial1}) are automatically satisfied and
the potential goes to zero at large distances from the guest charges, too.

\subsection{2D} \label{Sec21}
In 2D, the relevant function is the modified Bessel function $K_0$
\cite{Gradshteyn} which fulfills the differential equation
\begin{equation}
\frac{{\rm d}^2 K_0(r)}{{\rm d}r^2} + \frac{1}{r}
\frac{{\rm d}K_0(r)}{{\rm d}r} - K_0(r) = 0 
\end{equation}
and exhibits the following small- and large-distance asymptotics
\begin{equation}
K_0(r) \mathop{\sim}_{r\to 0} - \ln r , \qquad
K_0(r) \mathop{\sim}_{r\to \infty} \sqrt{\frac{\pi}{2 r}} {\rm e}^{-r} .  
\end{equation}  
In terms of $K_0$, the electric potential $\phi$ induced by the guest
charge $q e$ placed at the origin ${\bf 0}$ reads as
\begin{equation}
\phi(r\vert \{ q e,{\bf 0}\}) = q e K_0(\kappa r) .
\end{equation}
As required, it fulfills the small-distance asymptotic (\ref{crucial1})
and vanishes at $r\to\infty$.
The corresponding charge density
\begin{equation}
\rho(r\vert \{ q e,{\bf 0}\}) = - \frac{q e}{2\pi} \kappa^2 K_0(\kappa r)
\end{equation}
satisfies the condition
\begin{equation}
\int_0^{\infty} {\rm d}r\, 2\pi r \rho(r\vert \{ q e,{\bf 0}\}) =
- q e \int_0^{\infty} {\rm d}t\, t K_0(t) = -q e   
\end{equation}
which states that the charge cloud induced in the jellium is opposite
to the charge of the guest particle, for the overall neutrality reasons.

\begin{figure}[tbp]
\centering
\includegraphics[clip,width=0.8\textwidth]{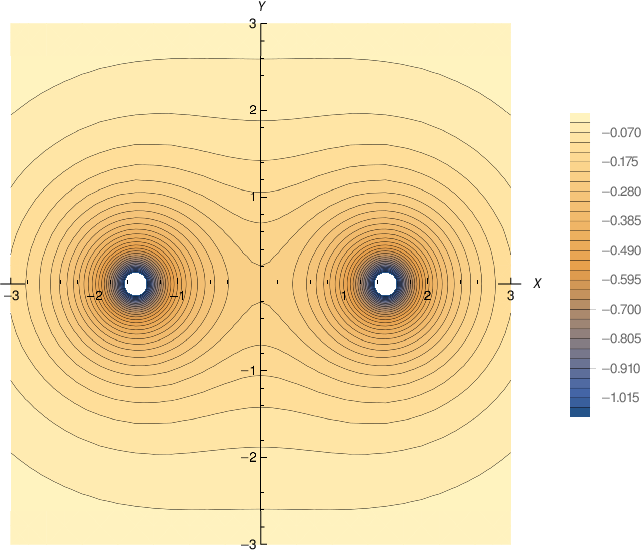}
\caption{The dimensionless charge density $\rho(x,y)/(e n_0)$ induced
by two pointlike charges with the valence $q_1=q_2=1$ at distance $d=3$,
fixed at the points ${\bf r}_1=(-3/2,0)$ and ${\bf r}_2=(3/2,0)$.
It is given by the Debye-H\"uckel formula (\ref{chargedens}),
in the units of the inverse Debye-H\"uckel length
$\kappa\equiv\sqrt{2\pi\beta e^2 n_0} = 1$ and $\pi n_0=1$.}
\label{fig1}
\end{figure}

Due to the superposition principle (\ref{superposition}), the charge density
at point ${\bf r}=(x,y)$, induced by two guest charges at a finite distance $d$,
the one with charge $q_1 e$ at point ${\bf r}_1=(-d/2,0)$
and the other with charge $q_2 e$ at point ${\bf r}_2=(d/2,0)$, reads as
\begin{eqnarray}
\rho\left({\bf r}\vert \{ q_1 e,(-d/2,0)\},\{ q_2 e,(d/2,0)\}\right) = 
\qquad \qquad \qquad \phantom{aaaa} \nonumber \\
- \frac{\kappa^2 e}{2\pi} \left[
q_1 K_0\left(\kappa\sqrt{\left(x+d/2\right)^2+y^2}\right)
+ q_2 K_0\left(\kappa\sqrt{\left(x-d/2\right)^2+y^2}\right) \right] .  
\label{chargedens}
\end{eqnarray}
The plot of the dimensionless charge density $\rho(x,y)/(e n_0)$ induced
by two unit charges with the valence $q_1=q_2=1$ at distance $d=3$,
fixed at the points ${\bf r}_1=(-3/2,0)$ and ${\bf r}_2=(3/2,0)$,
is pictured in figure \ref{fig1}. 

In polar coordinates $(r,\varphi)$ defined by $x=r\cos\varphi$ and
$y=r\sin\varphi$, the induced charge density at large distances from
the guest charges behaves as
\begin{equation} \label{separation}
\rho(r,\varphi) \mathop{\sim}_{r\to\infty} - \frac{\kappa^2 e}{2\pi}
\sqrt{\frac{\pi}{2\kappa r}} {\rm e}^{-\kappa r} \left[
q_1 {\rm e}^{-\frac{\kappa d}{2}\cos\varphi} + q_2 {\rm e}^{\frac{\kappa d}{2}\cos\varphi}
\right] .    
\end{equation}
The special feature of this asymptotic formula is its separation form in
distance $r$ and the angle $\varphi$, visible clearly in figure \ref{fig1}. 

\subsection{3D} \label{Sec22}
In 3D, the solution of the $\nu=3$ equation (\ref{crucial}), which
fulfills the small-distance asymptotic (\ref{crucial1}) and vanishes
at infinity, is of the Yukawa form
\begin{equation}
\phi(r\vert \{ q e,{\bf 0}\}) = q e \frac{\exp(-\kappa r)}{r} .
\end{equation}
The corresponding charge density induced in the jellium
\begin{equation}
\rho(r\vert \{ q e,{\bf 0}\}) = - \frac{q e}{4\pi} \kappa^2
\frac{\exp(-\kappa r)}{r}
\end{equation}
fulfills the electroneutrality condition
\begin{equation}
\int_0^{\infty} {\rm d}r\, 4\pi r^2 \rho(r\vert \{ q e,{\bf 0}\}) =
- q e \int_0^{\infty} {\rm d}t\, t {\rm e}^{-t} =  -q e .   
\end{equation}

In the case of two guest charges at a finite distance $d$, the one with
charge $q_1 e$ at point ${\bf r}_1=(-d/2,0,0)$ and the other with charge
$q_2 e$ at point ${\bf r}_2=(d/2,0,0)$, the charge density is given by
\begin{eqnarray}
\rho\left({\bf r}\vert \{ q_1 e,(-d/2,0,0)\},\{ q_2 e,(d/2,0,0)\}\right) = 
\qquad \qquad \phantom{aaaa} \nonumber \\ \qquad \qquad \qquad
- \frac{\kappa^2 e}{4\pi} \left[
q_1 \frac{\exp\left(-\kappa\sqrt{\left(x+d/2\right)^2+y^2+z^2}\right)}{
\sqrt{\left(x+d/2\right)^2+y^2+z^2}}  \right. \nonumber \\ \left.
\qquad \qquad \qquad \qquad \qquad
+ q_2 \frac{\exp\left(-\kappa\sqrt{\left(x-d/2\right)^2+y^2+z^2}\right)}{
\sqrt{\left(x-d/2\right)^2+y^2+z^2}} \right] .   
\end{eqnarray}
In spherical coordinates $(r,\varphi,\vartheta)$ defined by
$x=r\cos\varphi\sin\vartheta$, $y=r\sin\varphi\sin\vartheta$ and
$z=r\cos\vartheta$, the induced charge density at large distances from
the guest charges behaves as
\begin{equation}
\rho(r,\varphi,\vartheta) \mathop{\sim}_{r\to\infty}
- \frac{\kappa^2 e}{4\pi} \frac{{\rm e}^{-\kappa r}}{r}
\left[ q_1 {\rm e}^{-\frac{\kappa d}{2}\cos\varphi\sin\vartheta}
+ q_2 {\rm e}^{\frac{\kappa d}{2}\cos\varphi\sin\vartheta} \right] .    
\end{equation}
In analogy with the 2D case, the prefactor to the exponential decay
depends on the spherical angles $\varphi$ and $\vartheta$.

\renewcommand{\theequation}{3.\arabic{equation}}
\setcounter{equation}{0}

\section{General formalism in 2D} \label{Sec3}
Let us consider a system of $N$ mobile (pointlike) particles of charge $e$
with pair logarithmic interactions, confined to a 2D domain $\Lambda$
of surface $\vert\Lambda\vert$.
A uniform background charge density $\rho_b=-e N/\vert\Lambda\vert$
is fixed on the domain's surface to ensure the global electroneutrality.
The system is considered to be in thermal equilibrium at the inverse
temperature $\beta=1/(k_{\rm B}T)$.
The electric potential induced by the neutralizing background charge
\begin{equation}
e \phi({\bf r}) = -\rho_b \int_{\Lambda} {\rm d}^2r'\, 
\ln\vert {\bf r}-{\bf r}'\vert
\end{equation}  
implies the one-body Boltzmann factor
\begin{equation} \label{wr}
w({\bf r}) = \exp\left[ - \Gamma \phi({\bf r}) \right] ,
\qquad {\bf r}\in \Lambda ,  
\end{equation}  
where $\Gamma=\beta e^2$ is the coupling constant.
The two-body Boltzmann factor of two particles at spatial positions
${\bf r}$ and ${\bf r}'$ reads as $\vert {\bf r}-{\bf r}' \vert^{\Gamma}$.

\subsection{Homogeneous jellium} \label{Sec31}
The partition function of the jellium system
\begin{equation} \label{partition}
Z_N[w] = \frac{1}{N!} \int_{\Lambda} \left[ \prod_{j=1}^N {\rm d}^2r_j\,
w({\bf r}_j) \right]
\prod_{(j<k)=1}^N \vert {\bf r}_j-{\bf r}_k\vert^{\Gamma}
\end{equation}
is the functional of $w({\bf r})$.

The particle density at point ${\bf r}\in\Lambda$ is defined by
\begin{eqnarray} 
n_N({\bf r}) & = & \left\langle \sum_{j=1}^N \delta({\bf r}-{\bf r}_j)
\right\rangle \nonumber \\ & = & \frac{w({\bf r})}{(N-1)!Z_N} 
\int_{\Lambda} \left[ \prod_{j=1}^{N-1} {\rm d}^2r_j\, w({\bf r}_j) \right]
\prod_{j=1}^{N-1} \vert {\bf r}-{\bf r}_j \vert^{\Gamma}
\prod_{(j<k)=1}^{N-1} \vert {\bf r}_j-{\bf r}_k \vert^{\Gamma} ,
\nonumber \\ & & \label{onebody1}
\end{eqnarray}
where $\langle\cdots\rangle$ denotes the statistical average over
the canonical ensemble.
It can be obtained in the standard way as the functional derivative
of the partition function,
\begin{equation} \label{onebody2}
n_N({\bf r}) = w({\bf r}) \frac{1}{Z_N[w]}
\frac{\delta Z_N[w]}{\delta w({\bf r})} .
\end{equation}

Analogously, the two-body density
\begin{eqnarray} 
n_N^{(2)}({\bf r},{\bf r}') & = & \left\langle \sum_{(j\ne k)=1}^N
\delta({\bf r}-{\bf r}_j) \delta({\bf r}'-{\bf r}_k)
\right\rangle \nonumber \\ & = & \frac{w({\bf r}) w({\bf r}')}{(N-2)!Z_N}
\vert {\bf r}-{\bf r}'\vert^{\Gamma} 
\int_{\Lambda} \left[ \prod_{j=1}^{N-2} {\rm d}^2r_j\, w({\bf r}_j) \right]
\nonumber \\ & & \times \prod_{j=1}^{N-2} \vert {\bf r}-{\bf r}_j \vert^{\Gamma}
\vert {\bf r}'-{\bf r}_j \vert^{\Gamma}
\prod_{(j<k)=1}^{N-2} \vert {\bf r}_j-{\bf r}_k \vert^{\Gamma} \label{twobody1}
\end{eqnarray}
can be calculated as
\begin{equation} \label{twobody2}
n_N^{(2)}({\bf r},{\bf r}') = w({\bf r}) w({\bf r}') \frac{1}{Z_N[w]}
\frac{\delta^2 Z_N[w]}{\delta w({\bf r}) \delta w({\bf r}')} ,
\end{equation}
the three-body density
\begin{eqnarray} 
n_N^{(3)}({\bf r},{\bf r}',{\bf r}'') & = & \left\langle \sum_{[(j\ne k)\ne l]=1}^N
\delta({\bf r}-{\bf r}_j) \delta({\bf r}'-{\bf r}_k) \delta({\bf r}''-{\bf r}_l)
\right\rangle \nonumber \\ & = &
\frac{w({\bf r}) w({\bf r}') w({\bf r}'')}{(N-3)!Z_N}
\vert {\bf r}-{\bf r}'\vert^{\Gamma} \vert {\bf r}-{\bf r}''\vert^{\Gamma}
\vert {\bf r}'-{\bf r}''\vert^{\Gamma} \nonumber \\ & & \times
\int_{\Lambda} \left[ \prod_{j=1}^{N-3} {\rm d}^2r_j\, w({\bf r}_j) \right]
\prod_{j=1}^{N-3} \vert {\bf r}-{\bf r}_j \vert^{\Gamma}
\vert {\bf r}'-{\bf r}_j \vert^{\Gamma} \vert {\bf r}''-{\bf r}_j\vert^{\Gamma}
\nonumber \\ & & \times
\prod_{(j<k)=1}^{N-2} \vert {\bf r}_j-{\bf r}_k \vert^{\Gamma} \label{threebody1}
\end{eqnarray}
as 
\begin{equation} \label{threebody2}
n_N^{(3)}({\bf r},{\bf r}') = w({\bf r}) w({\bf r}') w({\bf r}'')
\frac{1}{Z_N[w]} \frac{\delta^3 Z_N[w]}{\delta
w({\bf r}) \delta w({\bf r}') \delta w({\bf r}'')} ,
\end{equation}
and so on.

\subsection{Jellium perturbed by one guest charge} \label{Sec32}
Let us perturb the jellium system by fixing a guest pointlike
particle of charge $e$ (equal to that of mobile plasma particles)
at a spatial position ${\bf r}'\in \Lambda$ and consider
the statistical quantities in its presence.
The partition function of $N-1$ mobile particles, in the presence of
the guest charge $e$ fixed at point ${\bf r}'$, is given by
\begin{eqnarray} 
Z_{N-1}(\{ e,{\bf r}'\}) & = & \frac{w({\bf r}')}{(N-1)!} 
\int_{\Lambda} \left[ \prod_{j=1}^{N-1} {\rm d}^2r_j\, w({\bf r}_j) \right]
\nonumber \\ & & \times \prod_{j=1}^{N-1} \vert {\bf r}'-{\bf r}_j \vert^{\Gamma}
\prod_{(j<k)=1}^{N-1} \vert {\bf r}_j-{\bf r}_k \vert^{\Gamma} . \label{partguest}
\end{eqnarray}
With regard to Eq. (\ref{onebody1}) this relation implies that
\begin{equation}
Z_{N-1}(\{ e,{\bf r}'\}) = Z_N n_N({\bf r}') ,
\end{equation}
where the rhs contains only unperturbed statistical quantities.
Thus the partition function in the presence of a guest charge
is expressible in terms of statistical quantities of the unperturbed jellium
system, namely the partition function and the particle density at the point
where the guest charge is placed.

The density of $N-1$ mobile particles at point ${\bf r}$, in the presence of
the guest charge $e$ fixed at ${\bf r}'$, is given by
\begin{eqnarray} 
n_{N-1}({\bf r}\vert\{ e,{\bf r}'\}) & = &
\left\langle \sum_{j=1}^{N-1} \delta({\bf r}-{\bf r}_j)
\right\rangle_{\{ e,{\bf r}'\}} \nonumber \\ & = &
\frac{w({\bf r}) w({\bf r}')}{(N-2)!Z_{N-1}(\{ e,{\bf r}'\})}
\vert {\bf r}-{\bf r}'\vert^{\Gamma} 
\int_{\Lambda} \left[ \prod_{j=1}^{N-2} {\rm d}^2r_j\, w({\bf r}_j) \right]
\nonumber \\ & & \times \prod_{j=1}^{N-2} \vert {\bf r}-{\bf r}_j \vert^{\Gamma}
\vert {\bf r}'-{\bf r}_j \vert^{\Gamma} \prod_{(j<k)=1}^{N-2}
\vert {\bf r}_j-{\bf r}_k \vert^{\Gamma} , \label{densityguest}
\end{eqnarray}
so that
\begin{equation}
n_{N-1}({\bf r}\vert\{ e,{\bf r}'\}) = \frac{Z_N}{Z_{N-1}(\{ e,{\bf r}'\})}
n_N^{(2)}({\bf r},{\bf r}') = \frac{n_N^{(2)}({\bf r},{\bf r}')}{
n_N({\bf r}')} .   
\end{equation}  

One can show in an analogous way that the two-body density of $N-1$ mobile
particles at points ${\bf r}_1$ and ${\bf r}_2$, in the presence of the guest
charge $e$ fixed at ${\bf r}'$, is given by
\begin{equation}
n^{(2)}_{N-1}({\bf r}_1,{\bf r}_2\vert\{ e,{\bf r}'\}) =
\frac{n_N^{(3)}({\bf r}_1,{\bf r}_2,{\bf r}')}{n_N({\bf r}')}    
\end{equation}
and so on.

\subsection{Jellium perturbed by two guest charges} \label{Sec33}
Perturbing the jellium system by two guest charges $e$, the one at point
${\bf r}'$ and the other at ${\bf r}''$, the partition function of $N-2$
mobile particles is expressible as
\begin{eqnarray} 
Z_{N-2}(\{ e,{\bf r}'\},\{ e,{\bf r}''\}) & = &
\frac{w({\bf r}') w({\bf r}'')}{(N-2)!} \vert {\bf r}'-{\bf r}''\vert^{\Gamma} 
\int_{\Lambda} \left[ \prod_{j=1}^{N-2} {\rm d}^2r_j\, w({\bf r}_j) \right]
\nonumber \\ & & \times \prod_{j=1}^{N-2} \vert {\bf r}'-{\bf r}_j\vert^{\Gamma}
\vert {\bf r}''-{\bf r}_j \vert^{\Gamma}
\prod_{(j<k)=1}^{N-2} \vert {\bf r}_j-{\bf r}_k \vert^{\Gamma} . \label{partguest2}
\end{eqnarray}
Consequently,
\begin{equation} \label{firstcrucial}
Z_{N-2}(\{ e,{\bf r}'\},\{ e,{\bf r}''\}) = Z_N n^{(2)}_N({\bf r}',{\bf r}'') .
\end{equation}

The density of $N-2$ mobile particles at point ${\bf r}$, in the presence of
the two guest charges, is given by
\begin{eqnarray} 
n_{N-2}({\bf r}\vert\{ e,{\bf r}'\},\{ e,{\bf r}''\}) & = &
\left\langle \sum_{j=1}^N \delta({\bf r}-{\bf r}_j)
\right\rangle_{\{ e,{\bf r}'\},\{ e,{\bf r}''\}} \nonumber \\ & = &
\frac{w({\bf r}) w({\bf r}') w({\bf r}'')}{(N-3)!
Z_{N-2}(\{ e,{\bf r}'\},\{ e,{\bf r}''\})}
\vert {\bf r}-{\bf r}'\vert^{\Gamma} \vert {\bf r}-{\bf r}''\vert^{\Gamma}
\nonumber \\ & & \times
\vert {\bf r}'-{\bf r}''\vert^{\Gamma} 
\int_{\Lambda} \left[ \prod_{j=1}^{N-3} {\rm d}^2r_j\, w({\bf r}_j) \right]
\prod_{j=1}^{N-3} \vert {\bf r}-{\bf r}_j \vert^{\Gamma} \nonumber \\ & & \times 
\vert {\bf r}'-{\bf r}_j \vert^{\Gamma} \vert {\bf r}''-{\bf r}_j \vert^{\Gamma}
\prod_{(j<k)=1}^{N-3} \vert {\bf r}_j-{\bf r}_k \vert^{\Gamma} .
\label{densityguest2}
\end{eqnarray}
Thus,
\begin{eqnarray}
n_{N-2}({\bf r}\vert\{ e,{\bf r}'\},\{ e,{\bf r}''\}) & = &
\frac{Z_N}{Z_{N-2}(\{ e,{\bf r}'\},\{ e,{\bf r}''\})}
n_N^{(3)}({\bf r},{\bf r}',{\bf r}'') \nonumber \\ & = &
\frac{n_N^{(3)}({\bf r},{\bf r}',{\bf r}'')}{n^{(2)}_N({\bf r}',{\bf r}'')} .   
\label{inducedcharge}
\end{eqnarray}  

Analogously, the two-body density of $N-2$ mobile particles at points
${\bf r}_1$ and ${\bf r}_2$, in the presence of the two guest
charges $e$, is given by
\begin{equation}
n^{(2)}_{N-2}({\bf r}_1,{\bf r}_2\vert\{ e,{\bf r}'\},\{ e,{\bf r}''\}) =
\frac{n_N^{(4)}({\bf r}_1,{\bf r}_2,{\bf r}',{\bf r}'')}{
n^{(2)}_N({\bf r}',{\bf r}'')}    
\end{equation}
and so on.

Note that the above formulas between the perturbed and unperturbed
statistical quantities are valid in any dimension, including 3D;
their derivation proceeds along the same lines.  

\renewcommand{\theequation}{4.\arabic{equation}}
\setcounter{equation}{0}

\section{The free-fermion coupling $\Gamma=2$} \label{Sec4}

\subsection{General formalism at the coupling $\Gamma=2$} \label{Sec41}
As was mentioned in the Introduction, the 2D jellium is exactly solvable
at the coupling $\Gamma=2$, in the bulk \cite{Jancovici81,Alastuey81},
on the surface of a sphere \cite{Caillol81}, on a semiperiodic strip
\cite{Choquard83}, in an inhomogeneous background \cite{Alastuey84},
with a variety of plane interfaces \cite{Cornu88}, etc. 

The coupling $\Gamma=2$ is the first of the series of the couplings
$\Gamma=2\gamma$ ($\gamma=1,2,3,\ldots$ a positive integer)
for which the partition function of the form (\ref{partition}) can be
expressed as an integral over grassman (anticommuting) numbers
\cite{Samaj95,Samaj04,Samaj15}.
In general, grassman algebra is generated by $M$ numbers $x_i$
$(i=1,\ldots,M)$ that anticommute, $\{ x_i,x_j \} := x_i x_j + x_j x_i =0$
\cite{Berezin66}; since $x_i^2=0$, such elements are nilpotent with degree 2.

In particular, for $\Gamma=2$ $(\gamma=1)$ one defines a discrete chain of
$N$ sites $j=0,1,\ldots,N-1$.
At each site $j$, there is one variable of type $\xi_j$ and one variable
of type $\psi_j$, all variables anticommute with each other, i.e.,
$\{ \xi_j,\xi_k \} = \{ \psi_j,\psi_k\} = \{ \xi_j,\psi_k\} = 0$
for all couples of sites $j,k=0,1,\ldots,N-1$. 
The multi-dimensional integral of the form (\ref{partition}) can be expressed
as an integral over anticommuting variables (for definition of integrals over
anticommuting variables, see \cite{Berezin66}) as follows \cite{Samaj95}
\begin{equation} \label{antipart}
Z_N[w] = \int {\cal D}\psi {\cal D}\xi\, {\rm e}^{S(\xi,\psi)} , 
\qquad S(\xi,\psi) = \sum_{j,k=0}^{N-1} \xi_j w_{jk} \psi_k .
\end{equation}
Here,
${\cal D}\psi {\cal D}\xi \equiv \prod_{j=0}^{N-1} {\rm d}\psi_j{\rm d}\xi_j$
and the action $S(\xi,\psi)$ involves pair interactions of anticommuting
variables via the overlap $N\times N$ matrix ${\bf w}$ whose elements
are given by
\begin{equation} \label{intmatrix}
w_{jk} = \int_{\Lambda} {\rm d}^2r\, w({\bf r}) z^j \bar{z}^k , \qquad
j,k=0,1,\ldots,N-1,  
\end{equation}  
where $z=x+{\rm i}y$ and $\bar{z}=x-{\rm i}y$ are complex coordinates
of the Cartesian point ${\bf r}=(x,y)$.
Because of the gaussian ``free-fermion'' action in (\ref{antipart}),
respecting the integration rules over anticommuting variables \cite{Berezin66},
the partition function is nothing but the determinant of the overlap
matrix:
\begin{equation}
Z_N[w] = {\rm Det}\, (w_{j,k}) \vert_{j,k=0}^{N-1} .
\end{equation}  

Having at one's disposal the anticommuting representation of the partition
function (\ref{antipart}), the particle density, defined by (\ref{onebody1})
and (\ref{onebody2}), reads as
\begin{equation} \label{density1}
n_N({\bf r}) = w({\bf r}) \sum_{j,k=0}^{N-1} \langle \xi_j \psi_k \rangle
z^j \bar{z}^k ,  
\end{equation}  
where
\begin{equation}
\langle \cdots \rangle = \frac{1}{Z_N[w]}
\int \prod_{j=0}^{N-1} \left[ {\rm d}\psi_j {\rm d}\xi_j \right]
{\rm e}^{S(\xi,\psi)} \cdots
\end{equation}
denotes the average over the anticommuting variables.
For our gaussian action $S(\xi,\psi)$, it holds that
\begin{equation} \label{density2}
\langle \xi_j \psi_k \rangle = w^{-1}_{kj} ,  
\end{equation}
where ${\bf w}^{-1}$ is the inverse of the overlap matrix ${\bf w}$.

The two-body density, defined by (\ref{twobody1}) and (\ref{twobody2}),
reads as
\begin{equation} \label{twobbody1}
n_N^{(2)}({\bf r}_1,{\bf r}_2) = w({\bf r}_1) w({\bf r}_2)
\sum_{j_1,k_1,j_2,k_2=0}^{N-1} \langle \xi_{j_1} \psi_{k_1} \xi_{j_2} \psi_{k_2}\rangle
z_1^{j_1} \bar{z}_1^{k_1} z_2^{j_2} \bar{z}_2^{k_2} .   
\end{equation}
The Wick theorem tells us that
\begin{equation} \label{twobbody2}
\langle \xi_{j_1} \psi_{k_1} \xi_{j_2} \psi_{k_2}\rangle =
\langle \xi_{j_1} \psi_{k_1} \rangle \langle \xi_{j_2} \psi_{k_2}\rangle -
\langle \xi_{j_1} \psi_{k_2} \rangle \langle \xi_{j_2} \psi_{k_1}\rangle .
\end{equation}

The three-body density, defined by (\ref{threebody1}) and (\ref{threebody2}),
is expressible as
\begin{eqnarray}
n_N^{(3)}({\bf r}_1,{\bf r}_2,{\bf r}_3) & = &
w({\bf r}_1) w({\bf r}_2) w({\bf r}_3) \sum_{j_1,k_1,j_2,k_2,j_3,k_3=0}^{N-1}
\nonumber \\ & & 
\langle \xi_{j_1} \psi_{k_1} \xi_{j_2} \psi_{k_2} \xi_{j_3} \psi_{k_3}\rangle
z_1^{j_1} \bar{z}_1^{k_1} z_2^{j_2} \bar{z}_2^{k_2} z_3^{j_3} \bar{z}_3^{k_3} .   
\end{eqnarray}
According to Wick's theorem,
\begin{eqnarray}
\langle \xi_{j_1} \psi_{k_1} \xi_{j_2} \psi_{k_2} \xi_{j_3} \psi_{k_3}\rangle & = &
\langle \xi_{j_1} \psi_{k_1} \rangle \langle \xi_{j_2} \psi_{k_2}\rangle
\langle \xi_{j_3} \psi_{k_3}\rangle -
\langle \xi_{j_1} \psi_{k_1} \rangle \langle \xi_{j_2} \psi_{k_3}\rangle
\langle \xi_{j_3} \psi_{k_2}\rangle \nonumber \\ & - &
\langle \xi_{j_1} \psi_{k_3} \rangle \langle \xi_{j_2} \psi_{k_2}\rangle
\langle \xi_{j_3} \psi_{k_1}\rangle -
\langle \xi_{j_1} \psi_{k_2} \rangle \langle \xi_{j_2} \psi_{k_1}\rangle
\langle \xi_{j_3} \psi_{k_3}\rangle \nonumber \\ & + &
\langle \xi_{j_1} \psi_{k_2} \rangle \langle \xi_{j_2} \psi_{k_3}\rangle
\langle \xi_{j_3} \psi_{k_1}\rangle +
\langle \xi_{j_1} \psi_{k_3} \rangle \langle \xi_{j_2} \psi_{k_1}\rangle
\langle \xi_{j_3} \psi_{k_2}\rangle . \nonumber \\ & &
\end{eqnarray}

\subsection{Circularly symmetric system} \label{Sec42}
We shall consider the jellium system with ``soft'' boundary, namely 
an infinite background charge density (with circular symmetry).
Increasing successively the finite number of particles $N$ to infinity,
we shall look for the particle density, the two-body density, etc. close
to the origin $r=0$, neglecting in this way the boundary effects
at large distances $r\propto \sqrt{N}$.
In the limit $N\to\infty$, the particle density $n(r)$ is expected to be
uniform for finite $r$, the two-body density $n^{(2)}({\bf r}_1,{\bf r}_2)$
is expected to be translationally invariant, i.e., dependent on the distance
between the points $\vert {\bf r}_1-{\bf r}_2 \vert$, etc.

An infinite uniform background charge density $\rho=-e n_0$ with the circular
symmetry induces the potential with the circular symmetry $\phi(r)$
given by the 2D angle-independent Poisson equation
\begin{equation}
\frac{{\rm d}^2\phi}{{\rm d}r^2} + \frac{1}{r}
\frac{{\rm d}\phi}{{\rm d}r} = 2\pi e n_0 . 
\end{equation}  
Up to an irrelevant constant, its solution reads as
\begin{equation}
\phi(r) = \frac{\pi}{2} e n_0 r^2 .
\end{equation}  
The corresponding one-body Boltzmann weight at the coupling
$\Gamma = \beta e^2 = 2$ is
\begin{equation}
w(r) = \exp\left( -\beta e \phi \right) = \exp\left( -\pi n_0 r^2 \right) .
\end{equation}  
With this Boltzmann weight, the overlap matrix (\ref{intmatrix})
becomes diagonal:
\begin{equation}
w_{jk} = \int_0^{\infty} {\rm d}r\, r^{1+j+k} \int_0^{2\pi} {\rm d}\varphi\,
\exp\left[ {\rm i}(j-k)\varphi \right]
= \frac{j!}{n_0 (\pi n_0)^j} \delta_{jk} .  
\end{equation}  
The inversion of the diagonal matrix is trivial,
\begin{equation}
w^{-1}_{jk} = \frac{n_0 (\pi n_0)^j}{j!} \delta_{jk} .  
\end{equation}  

For a finite number of particles $N$, the particle density, given by
(\ref{density1}) and (\ref{density2}), reads as
\begin{equation}
n_N(r) = {\rm e}^{-\pi n_0 r^2} n_0 \sum_{j=0}^{N-1}
\frac{(\pi n_0 r^2)^j}{j!} . 
\end{equation}
In the thermodynamic limit $N\to\infty$, the particle density in the bulk
interior of the jellium becomes uniform,
$\lim_{N\to\infty} n_N(r) \equiv n(r)=n_0$, and equal to the background charge
density (divided by $e$) to maintain the overall charge neutrality.

The two-body density, given by (\ref{twobbody1}) and (\ref{twobbody2}),
reads as
\begin{eqnarray}
n_N^{(2)}({\bf r}_1,{\bf r}_2) & = & n_N({\bf r}_1) n_N({\bf r}_2)
- {\rm e}^{-\pi n_0(r_1^2+r_2^2)} n_0^2
\sum_{j=0}^{N-1} \frac{(\pi n_0 z_1 \bar{z}_2)^j}{j!}
\nonumber \\ & & \times
\sum_{k=0}^{N-1} \frac{(\pi n_0 \bar{z}_1 z_2)^k}{k!} .
\end{eqnarray}
In the limit $N\to\infty$, since
$(z_1-z_2)(\bar{z}_1-\bar{z}_2) = \vert {\bf r}_1-{\bf r}_2\vert^2$
one obtains that
\begin{equation} \label{twothermo}
\frac{n^{(2)}({\bf r}_1,{\bf r}_2)}{n_0^2} = 1 -
\exp\left( -\pi n_0 \vert {\bf r}_1-{\bf r}_2\vert^2 \right) ,  
\end{equation}
i.e., the two-body density is translation-invariant as was expected.

The three-body density, given by (\ref{threebody1}) and (\ref{threebody1}),
takes in the limit $N\to\infty$ the form
\begin{eqnarray}
\frac{n^{(3)}({\bf r}_1,{\bf r}_2,{\bf r}_3)}{n_0^3} & = & 1 -
\exp\left( -\pi n_0 \vert {\bf r}_1-{\bf r}_2\vert^2 \right) \nonumber \\
& & - \exp\left( -\pi n_0 \vert {\bf r}_1-{\bf r}_3\vert^2 \right) 
- \exp\left( -\pi n_0 \vert {\bf r}_2-{\bf r}_3\vert^2 \right) \nonumber \\
& & + \exp\left[ -\pi n_0 \left( z_1\bar{z}_1 + z_2\bar{z}_2 + z_3\bar{z}_3
- z_1\bar{z}_2 - z_2\bar{z}_3 - z_3\bar{z}_1 \right) \right] \nonumber \\
& & + \exp\left[ -\pi n_0 \left( z_1\bar{z}_1 + z_2\bar{z}_2 + z_3\bar{z}_3
- z_1\bar{z}_3 - z_2\bar{z}_1 - z_3\bar{z}_2 \right) \right] . \nonumber \\
& & \label{three}
\end{eqnarray}
It can be checked that this expression is invariant with respect to
a uniform shift of all particle coordinates $z_j\to z_j+\delta$,
$\bar{z}_j\to {\bar z}_j+\bar{\delta}$, as it should be. 

\renewcommand{\theequation}{5.\arabic{equation}}
\setcounter{equation}{0}

\section{Two identical unit guest charges at $\Gamma=2$} \label{Sec5}
In this part, we consider the problem of two identical pointlike particles
with the elementary charge $e$ at distance $d$, immersed in the bulk interior
of the 2D jellium at the free-fermion coupling $\Gamma=2$.
Let one of the guest charges be fixed at point ${\bf r}'=(-d/2,0)$ and
the other one at point ${\bf r}''=(d/2,0)$.
The formalism developed above is applied to the calculation of the effective
force between the guest charges and the induced charge density in the jellium.

\subsection{Effective force between guest charges} \label{Sec51}
The free energy of $N-2$ particles plus the two considered guest charges
reads as
\begin{eqnarray}
- \beta F_{N-2}(\{ e,{\bf r}'\},\{ e,{\bf r}''\}) & = &
\ln Z_{N-2}(\{ e,{\bf r}'\},\{ e,{\bf r}''\}) \nonumber \\ & = &
\ln Z_N + \ln n^{(2)}_N({\bf r}',{\bf r}'') , \label{eq1}
\end{eqnarray}
where the formula (\ref{firstcrucial}) was taken into account.
The corresponding force between the two guest charges at distance $d$,
${\cal F}_{N-2}(d)$, is given by
\begin{equation} \label{force}
{\cal F}_{N-2}(d) = - \frac{\partial F_{N-2}}{\partial d} .
\end{equation}
A positive/negative sign of the force means an effective repulsion/attraction
between the guest charges. 
Since $\ln Z_N$ on the rhs of (\ref{eq1}) does not depend on $d$,
the formula for the force (\ref{force}) results in
\begin{equation}
\beta {\cal F}_{N-2}(d) = \frac{\partial}{\partial d}
\ln n^{(2)}_N({\bf r}',{\bf r}'') .
\end{equation}
In the thermodynamic limit $N\to\infty$, taking the expression
(\ref{twothermo}) for the two-body density and considering that
$\beta e^2 = 2$, one gets
\begin{equation} \label{forceresult}
\frac{{\cal F}(d)}{e^2} = \pi n_0 d \frac{1}{{\rm e}^{\pi n_0 d^2}-1} . 
\end{equation}
This means that ${\cal F}(d)>0$ for an arbitrary finite distance $d$ between
the two guest charges, i.e., the force is repulsive. 
Its small-distance behavior
\begin{equation}
\lim_{d\to 0} {\cal F}(d) \sim \frac{e^2}{d}
\end{equation}
can be explained by the fact that the pure Coulomb interaction $- e^2 \ln d$
is relevant for small distance between the guest charges
(i.e., the interaction with the jellium particles can be neglected), so that
${\cal F}(d) \sim - (-e^2) \partial \ln d/\partial d$.
At large distance between the guest charges, it holds that
\begin{equation}
\lim_{d\to\infty} {\cal F}(d) \sim e^2 \pi n_0 d {\rm e}^{-\pi n_0 d^2} , 
\end{equation}
i.e., the decay of the repulsive force to zero is short-ranged
of Gaussian type.

\begin{figure}[tbp]
\centering
\includegraphics[clip,width=0.8\textwidth]{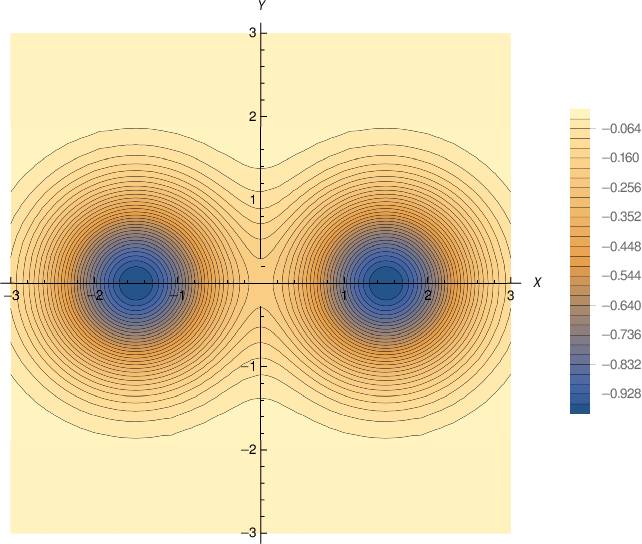}
\caption{The dimensionless charge density $\rho(x,y)/(e n_0)$ induced
by two pointlike charges with the valence $q_1=q_2=1$ at distance $d=3$,
fixed at the points ${\bf r}_1=(-3/2,0)$ and ${\bf r}_2=(3/2,0)$.
It is given by the formula (\ref{cloud}) valid for the jellium coupling
$\Gamma=2$, in the units of $\pi n_0=1$.}
\label{fig2}
\end{figure}

\subsection{Charge cloud induced in jellium by the guest particles} \label{Sec52}
The excess charge density, induced in the jellium due to the presence of
the two guest charges, is obtained by using Eq. (\ref{inducedcharge})
for the points ${\bf r}=(x,y)$, ${\bf r}'=(-d/2,0)$ and ${\bf r}''=(d/2,0)$.
With the aid of formula (\ref{three}) for the three-body density of the
homogenous jellium, after simple algebra one arrives at
\begin{eqnarray}
\frac{\rho({\bf r}\vert\{ e,{\bf r}'\},\{ e,{\bf r}''\})}{e n_0} & = &
- \frac{{\rm e}^{-\pi n_0r^2}}{\sinh(\pi n_0 d^2/2)} \left[
{\rm e}^{\pi n_0d^2/4} \cosh(\pi n_0 d x) \right. \nonumber \\ & & \left. \qquad 
- {\rm e}^{-\pi n_0d^2/4} \cos(\pi n_0 d y) \right] . \label{cloud}  
\end{eqnarray}
The dimensionless charge density $\rho(x,y)/(e n_0)$ at the jellium coupling
$\Gamma=2$, induced by two unit charges at distance $d=3$ fixed at
the points ${\bf r}_1=(-3/2,0)$ and ${\bf r}_2=(3/2,0)$, is pictured
in the units of $\pi n_0=1$ in figure \ref{fig2}.
Note that in comparison with the Debye-H\"uckel contour plot in
figure \ref{fig1} the depletion region of plasma particles is more
concentrated in the neighbourhood of the guest particles.

Since $x=r\cos\varphi$ and $y=r\sin\varphi$, the formula (\ref{cloud}) has not
the simple separation Debye-H\"{u}ckel form (\ref{separation}).
It is likely that the separation of distance and angle variables
in the asymptotic large-distance formula for the excess charge density
is restricted to the high-temperature Debye-H\"uckel limit only.

\renewcommand{\theequation}{6.\arabic{equation}}
\setcounter{equation}{0}

\section{Effective interaction between two distinct guest charges} \label{Sec6}

\subsection{Pointlike charges $q e$ ($q=1,2,\ldots$) and $e$} \label{Sec61}
Let us now consider the case of two pointlike guest particles at distance $d$,
the one with charge $q e$ (the valence $q=1,2,\ldots$) and the other with
the elementary charge $e$.
For reasons which will be clear later, the guest charge $q e$ is placed at
the symmetry origin of the background charge ${\bf 0}$ and the charge
$e$ is placed at the radial distance $r=d$ from the origin ${\bf 0}$.

Adopting the notation $w({\bf r})=\exp(-\pi n_0r^2)$,
in analogy with the relation (\ref{partguest2}) it holds that
\begin{eqnarray} 
Z_{N-1}(\{ qe,{\bf 0}\},\{ e,{\bf r}\}) & = &
\frac{w({\bf r})}{(N-1)!} r^{\Gamma q} 
\int_{\Lambda} \left[ \prod_{j=1}^{N-1} {\rm d}^2r_j\, w({\bf r}_j) \right]
\nonumber \\ & & \times \prod_{j=1}^{N-1} r_j^{\Gamma q}
\vert {\bf r}-{\bf r}_j \vert^{\Gamma}
\prod_{(j<k)=1}^{N-1} \vert {\bf r}_j-{\bf r}_k \vert^{\Gamma} . \label{partguest3}
\end{eqnarray}
Similarly,
\begin{eqnarray} 
n_N({\bf r}\vert \{ qe,{\bf 0}\}) & = &
\frac{w({\bf r})}{(N-1)! Z_N(\{ qe,{\bf 0}\})} r^{\Gamma q} 
\int_{\Lambda} \left[ \prod_{j=1}^{N-1} {\rm d}^2r_j\, w({\bf r}_j) \right]
\nonumber \\ & & \times \prod_{j=1}^{N-1} r_j^{\Gamma q}
\vert {\bf r}-{\bf r}_j \vert^{\Gamma}
\prod_{(j<k)=1}^{N-1} \vert {\bf r}_j-{\bf r}_k \vert^{\Gamma} . 
\end{eqnarray}
Consequently,
\begin{equation}
Z_{N-1}(\{ qe,{\bf 0}\},\{ e,{\bf r}\}) =
Z_N(\{ qe,{\bf 0}\}) n_N({\bf r}\vert \{ qe,{\bf 0}\}) .  
\end{equation}
The corresponding free energy is given by
\begin{eqnarray}
- \beta F_{N-1}(\{ qe,{\bf 0}\},\{ e,{\bf r}\}) & = &
\ln Z_{N-1}(\{ qe,{\bf 0}\},\{ e,{\bf r}\}) \nonumber \\ & = &
\ln Z_N(\{ qe,{\bf 0}\})
+ \ln n_N(r\vert \{ qe,{\bf 0}\}) . \label{eq3}
\end{eqnarray}
Setting $r=d$, taking the thermodynamic limit $N\to\infty$ and noting
that $\ln Z_N(\{ qe,{\bf 0}\})$ on the rhs of (\ref{eq3}) does not depend
on $d$, the formula for the force (\ref{force}) results in
\begin{equation} \label{force1}
\beta {\cal F}(d,q) = \frac{\partial}{\partial d}
\ln n(d\vert \{ qe,{\bf 0}\}) .
\end{equation}

Fixing the guest charge $q e$ at the origin ${\bf 0}$, the one-body
Boltzmann weight $w({\bf r})=\exp(-\pi n_0r^2)$ is replaced by
\begin{equation}
\widetilde{w}({\bf r}) = r^{\Gamma q} \exp(-\pi n_0r^2) .
\end{equation}  
This Boltzmann weight still exhibits the radial symmetry and implies
the diagonal form of the overlap matrix (\ref{intmatrix}),
$\widetilde{w}_{jk} = \widetilde{w}_j \delta_{jk}$.
In particular, at the free-fermion coupling $\Gamma=2$,
\begin{equation}
\widetilde{w}_j = 2\pi \int_0^{\infty} {\rm d}r\, r r^{2(j+q)} {\rm e}^{-\pi n_0r^2}
= \frac{(j+q)!}{n_0 (\pi n_0)^{j+q}} , \quad j=0,1,\ldots,N-1.  
\end{equation}  
The particle density of interest
\begin{equation}
n_N(r\vert \{ qe,{\bf 0}\}) = \widetilde{w}(r) \sum_{j=0}^{N-1}
\frac{r^{2j}}{\widetilde{w}_j}  
\end{equation}
takes, in the thermodynamic limit $N\to\infty$, the form
\begin{equation}
\frac{n(r\vert \{ qe,{\bf 0}\})}{n_0} = 1 -
\exp(-\pi n_0r^2) \sum_{j=0}^{q-1} \frac{(\pi n_0 r^2)^j}{j!} .  
\end{equation}
It is simple to verify the electroneutrality sum rule
\begin{equation}
e \int_0^{\infty} {\rm d}^2r\, \left[ n(r\vert \{ qe,{\bf 0}\}) - n_0 \right]
= - q e  
\end{equation}  
which states that the charge induced in the plasma is opposite to that
of the guest charge.

Noting that we consider $\beta e^2=2$, the effective force (\ref{force1})
between the guest charges $q e$ and $e$ at distance $d$ in the jellium bulk
interior is given by
\begin{equation} \label{forceqe}
\frac{{\cal F}(d,q)}{e^2} = \pi n_0 d \frac{(\pi n_0 d^2)^{q-1}}{(q-1)!}
\frac{1}{{\rm e}^{\pi n_0 d^2}-\sum_{j=0}^{q-1}\frac{(\pi n_0 d^2)^j}{j!}} . 
\end{equation}
For $q=1$, one recovers the previous result (\ref{forceresult})
for the force between two elementary charges $e$ at distance $d$.
It is seen that the effective force is always positive (i.e., repulsive),
for any valence $q=1,2,\ldots$ of the guest charge and any distance $d$
between the guest charges.
The small-distance asymptotic behavior
\begin{equation}
{\cal F}(d,q) \mathop{\sim}_{d\to 0} \frac{q e^2}{d}
\end{equation}
comes from the pure Coulomb interaction between the guest charges
$- q e^2 \ln d$.
At large distance between the guest charges,
\begin{equation}
{\cal F}(d,q) \mathop{\sim}_{d\to\infty} e^2 \pi n_0 d
\frac{(\pi n_0 d^2)^{q-1}}{(q-1)!} {\rm e}^{-\pi n_0 d^2} , 
\end{equation}
the decay of the force to zero is Gaussian.

\begin{figure}[tbp]
\centering
\includegraphics[clip,width=0.6\textwidth]{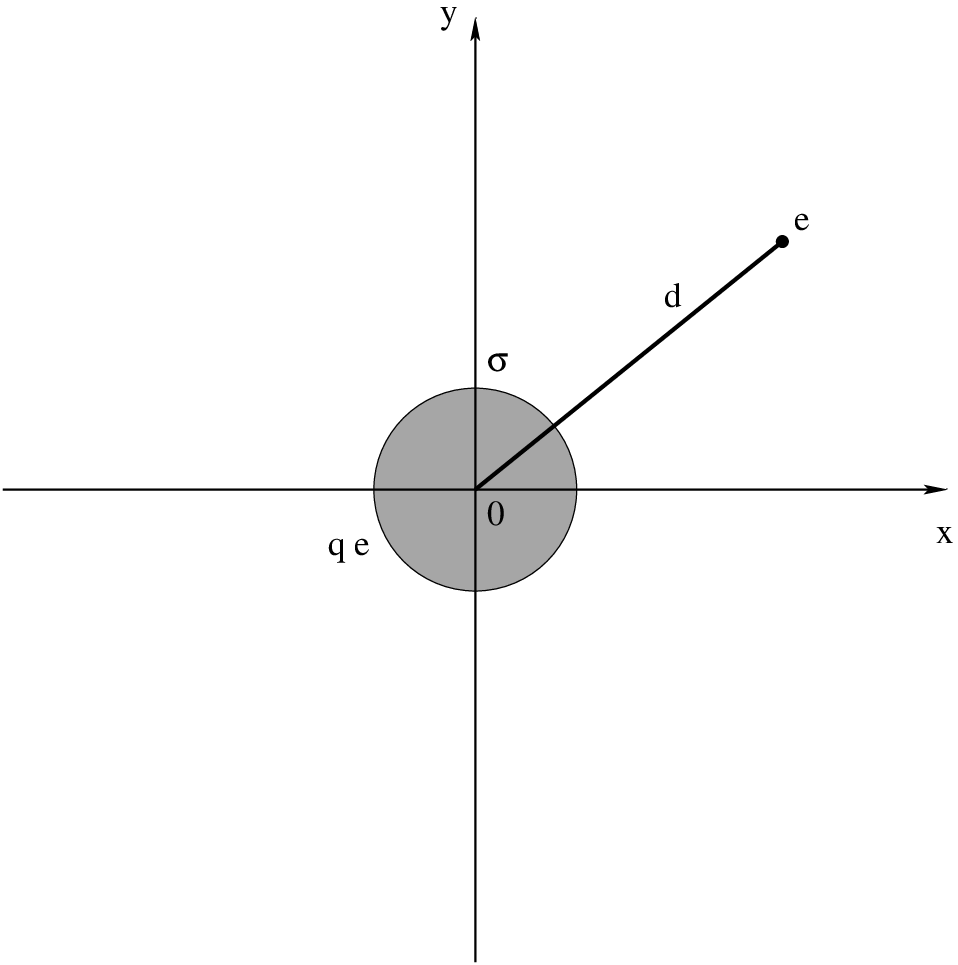}
\caption{The geometry of one guest particle with the charge
$q e$ ($q=0,\pm 1,\pm 2,\ldots$) and a hard core of radius $\sigma$
(shaded region), put at the origin ${\bf 0}$, and the other pointlike
one with charge $e$ at distance $d$ from the origin.}
\label{fig3}
\end{figure}

\subsection{Charge $q e$ with a hard core of radius $\sigma$ and pointlike
  charge $e$} \label{Sec62}
Let us now consider the geometry of the guest particle with the charge
$q e$ ($q=0,\pm 1,\pm 2,\ldots$) and a hard core of radius $\sigma$,
put at the origin ${\bf 0}$, and the pointlike guest charge $e$ at distance $d$
from the origin (see figure \ref{fig3}).
Although this model is slightly different from the previous one treated in
\cite{Ma01}, it involves all relevant physical phenomena of the original
problem.
The case $q=0$ corresponds to an empty void of radius $\sigma$;
notice that an empty void is not neutral due to the presence of the fixed
negative background charge density inside the hard core.
The presence of the hard core of radius $\sigma>0$ is crucial for negative
values of $q=-1,-2,\ldots$.
For two pointlike particles of charge $e$ and $q e$ at distance $r$,
the corresponding Boltzmann factor $r^{\beta q e^2}=r^{\Gamma q}$ is integrable
at small distances in 2D provided that $\Gamma q>-2$.
This means that for the coupling $\Gamma=2$ the guest particle with
$q=-1,-2,\ldots$ collapses thermodynamically with an arbitrary
particle of the plasma.
The collapse is avoided when the guest particle has a hard core which
prevents a contact with jellium particles.

At the free-fermion coupling $\Gamma=2$, the overlap matrix (\ref{intmatrix})
is diagonal with the elements
\begin{equation}
\widetilde{w}_j = 2\pi \int_{\sigma}^{\infty} {\rm d}r\, r r^{2(j+q)}
{\rm e}^{-\pi n_0r^2}
= \frac{1}{n_0 (\pi n_0)^{j+q}} \Gamma(1+j+q,\pi n_0\sigma^2) ,
\end{equation}
$j=0,1,\ldots,N-1$, where $\Gamma(a,x)$ is the incomplete Gamma function
defined by
\begin{equation} \label{incompleteGamma}
\Gamma(a,x) = \int_x^{\infty} {\rm d}t\, t^{a-1} {\rm e}^{-t} .
\end{equation}  
It satisfies the following relations \cite{Gradshteyn}
\begin{equation} \label{relation1}
\Gamma(a+1,x) = a \Gamma(a,x) + x^a {\rm e}^{-x} , 
\end{equation}
\begin{equation} \label{relation2}
\Gamma(a,x) =  \Gamma(a) - \sum_{k=0}^{\infty} (-1)^k
\frac{x^{a+k}}{k! (a+k)} ,
\end{equation}
\begin{equation} \label{relation3}
\Gamma(n,x) = \Gamma(n) {\rm e}^{-x} \sum_{k=0}^{n-1} \frac{x^k}{k!} ,
\qquad n=1,2,\ldots , 
\end{equation}
where
\begin{equation} \label{Gamma}
\Gamma(a) = \int_0^{\infty} {\rm d}t\, t^{a-1} {\rm e}^{-t} 
\end{equation}  
is the standard Gamma function.
The thermodynamic $N\to\infty$ limit of the particle density, defined for
the distance from the origin $r\ge\sigma$, is given by
\begin{equation} \label{densityg1}
\frac{n(r\vert \{ qe,\sigma,{\bf 0}\})}{n_0} =
{\rm e}^{-\pi n_0r^2} \sum_{j=0}^{\infty}
\frac{(\pi n_0 r^2)^{(j+q)}}{\Gamma(1+j+q,\pi n_0\sigma^2)} .
\end{equation}
The force (\ref{force1}) between the guest charges $q e$ with the hard core of
radius $\sigma$ and the pointlike charge $e$ at distance $d\ge \sigma$
reads as
\begin{equation} \label{forceg1}
\frac{{\cal F}(d,q)}{e^2} = -\pi n_0 d + \frac{1}{d} \frac{1}{\sum_{j=0}^{\infty}
\frac{(\pi n_0 d^2)^{j+q}}{\Gamma(1+j+q,\pi n_o\sigma^2)}} \sum_{j=0}^{\infty}
\frac{(j+q) (\pi n_0 d^2)^{j+q}}{\Gamma(1+j+q,\pi n_o\sigma^2)} .   
\end{equation}

To simplify the notation, we shall work in the units of $\pi n_0 = 1$.
In terms of the dimensionless distance $\sqrt{\pi n_0} r\to r$ and hard-core
radius $\sqrt{\pi n_0} \sigma\to \sigma$, the expression for the particle
density (\ref{densityg1}) takes the form
\begin{eqnarray} 
\frac{n(r\vert \{ qe,\sigma,{\bf 0}\})}{\pi n_0} & = &
\frac{1}{\pi} {\rm e}^{-r^2} \sum_{j=0}^{\infty}
\frac{r^{2(j+q)}}{\Gamma(1+j+q,\sigma^2)} \nonumber \\
& = & \frac{1}{\pi} \left[
1 - {\rm e}^{-r^2} \sum_{j=0}^{q-1} \frac{r^{2j}}{j!} \right] \label{densityg2}
\end{eqnarray}
and the expression for the dimensionless force between the guest charges
(\ref{forceg1}), ${\cal F}(d,q)/(e^2 \sqrt{\pi n_0})\to {\cal F}(d,q)$
takes the form
\begin{eqnarray} 
{\cal F}(d,q) & = & -d + \frac{1}{d}
\frac{1}{\sum_{j=0}^{\infty} \frac{d^{2(j+q)}}{\Gamma(1+j+q,\sigma^2)}}
\sum_{j=0}^{\infty} \frac{(j+q) d^{2(j+q)}}{\Gamma(1+j+q,\sigma^2)} \nonumber \\
& = & \frac{1}{d \sum_{j=0}^{\infty} \frac{d^{2(j+q)}}{\Gamma(1+j+q,\sigma^2)}}
\Bigg[ \frac{q d^{2q}}{\Gamma(1+q,\sigma^2)} \nonumber \\ & &
\qquad - {\rm e}^{-\sigma^2}
\sum_{j=1}^{\infty} \frac{(\sigma d)^{2(j+q)}}{\Gamma(j+q,\sigma^2)
\Gamma(1+j+q,\sigma^2)} \Bigg] . \label{forceg2}
\end{eqnarray}
Here, the formula (\ref{relation1}) was applied.

To analyse the last two relations, one has to distinguish between
the non-positive values of $q=0, -1, -2,\ldots$ and positive $q=1,2,\ldots$.

\subsubsection{$q=0,-1,-2,\ldots$}
In the case of the empty void with $q=0$ or a negatively charged guest
particle $q=-1,-2,\ldots$, since the incomplete Gamma functions in
(\ref{forceg2}) are positive it follows that
\begin{equation}
{\cal F}(d,q) < 0 ,
\end{equation}
i.e., the effective force between oppositely charged guest particle is
always attractive, for an arbitrary distance $d$ between them.
This is in agreement with one's intuition. 
It is questionable whether this result, valid at the coupling $\Gamma=2$,
holds also for higher values of $\Gamma$.
We suggest that stronger polarisation effects of positively charged particles
close to the negatively charged void will probably reduce the effective
attractive force between the guest particles which might even change
the sign of this force in certain ranges of the inter-particle distance $d$.

\subsubsection{$q=1, 2,\ldots$}
In the case of positively charged guest particle $q=1,2,\ldots$, using
the relation (\ref{relation3}) for the incomplete Gamma function
the expression for the particle density (\ref{densityg2}) can be written as
\begin{eqnarray}
\frac{n(r)}{\pi n_0} & = & \frac{1}{\pi} {\rm e}^{-r^2}
\sum_{j=0}^{\infty} \frac{r^{2(j+q)}}{(j+q)!}
\frac{\exp(\sigma^2)}{\sum_{k=0}^{j+q} \sigma^{2k}/k!} \nonumber \\
& = & \frac{1}{\pi} {\rm e}^{-r^2}
\sum_{j=0}^{\infty} \frac{r^{2(j+q)}}{(j+q)!}
\left[ 1 + \frac{\sum_{k=j+q+1}^{\infty} \sigma^{2k}/k!}{
\sum_{k=0}^{j+q} \sigma^{2k}/k!} \right] . \label{partdens}
\end{eqnarray}
In the limit of small hard core $\sigma$, the function in the square bracket
can be expanded in powers of $\sigma$ as follows
\begin{equation}
1 + \frac{\sigma^{2(j+q+1)}}{(j+q+1)!}
- \frac{\sigma^{2(j+q+2)}}{(j+q)!(j+q+2)} + \cdots .
\end{equation}
Based on this expansion, the formula (\ref{partdens}) can be rewritten as
\begin{equation} \label{finala}
\frac{n(r)}{\pi n_0} = \frac{1}{\pi} \left\{
1 - {\rm e}^{-r^2} \sum_{j=0}^{q-1} \frac{r^{2j}}{j!} +
{\rm e}^{-r^2} \sum_{j=q}^{\infty} \frac{\sigma^{2(j+1)}r^{2j}}{j!(j+1)!}
\left[ 1 + O\left( \frac{1}{r^2} \right) \right] \right\} .
\end{equation}
Two clearly separated contributions to the particle density are seen:
the one coming from the pure Coulomb interaction of the pointlike $q e$
and $e$ charges, see Eq. (\ref{densityg2}), and the other one
due to the hard core of charge $q e$ with radius $\sigma$. 
The important feature is that, in the large-$r$ limit, the subdominant term
of the pure Coulomb part $\propto {\rm e}^{-r^2} r^{2(q-1)}$ goes to zero more
quickly than the asymptotic terms attached to the even powers of $\sigma$.
Neglecting terms of order $O(1/r^2)$ in the square brackets
on the rhs of (\ref{finala}), in the large-$r$ limit one arrives at
\begin{equation}
\frac{n(r)}{\pi n_0} \mathop{\sim}_{r\to\infty} \frac{1}{\pi} \left\{
1 - {\rm e}^{-r^2} \sum_{j=0}^{q-1} \frac{r^{2j}}{j!} 
+ \sigma^2 {\rm e}^{-r^2} \left[ \frac{I_1(2\sigma r)}{\sigma r}
- \sum_{j=0}^{q-1} \frac{(\sigma r)^{2j}}{j!(j+1)!} \right] \right\} ,
\end{equation}
where
\begin{equation}
I_1(z) = \sum_{j=0}^{\infty} \frac{1}{j!(j+1)!}
\left( \frac{z}{2} \right)^{2j+1}   
\end{equation}
is the modified Bessel functions of the first kind \cite{Gradshteyn}.
Since
\begin{equation}
I_{\alpha}(z) \mathop{\sim}_{z\to\infty} \frac{{\rm e}^z}{\sqrt{2\pi z}} ,
\end{equation}
for a finite valence $q$ of the guest charge it holds that
\begin{equation}
\frac{n(r)}{\pi n_0} \mathop{\sim}_{r\to\infty} \frac{1}{\pi} \left[
1 + \frac{\sigma^2}{2\sqrt{\pi}} {\rm e}^{-r^2}
\frac{{\rm e}^{2\sigma r}}{(\sigma r)^{3/2}} \right] .
\end{equation}
Using formula (\ref{force1}), the effective force between the two charges
is given by
\begin{equation} \label{forceasym}
{\cal F}(d,q) \mathop{\sim}_{d\to\infty}
- \sqrt{\frac{\sigma}{\pi d}} {\rm e}^{-d^2+2\sigma d} .
\end{equation}
As is expected, the effective force vanishes for an infinite distance
since two guest charges do not feel one another when $d\to\infty$.
It goes to $0$ from below at asymptotically large (finite) distances $d$,
i.e., it is attractive.
In other words, at asymptotically large (finite) distances $d$,
the attractive interaction of the guest charge $e$ with the void
due to the hard core of the guest particle $q e$ gets over
the repulsive (effective) Coulomb interaction between these guest charges. 

\begin{figure}[tbp]
\centering
\includegraphics[clip,angle=-90,width=0.8\textwidth]{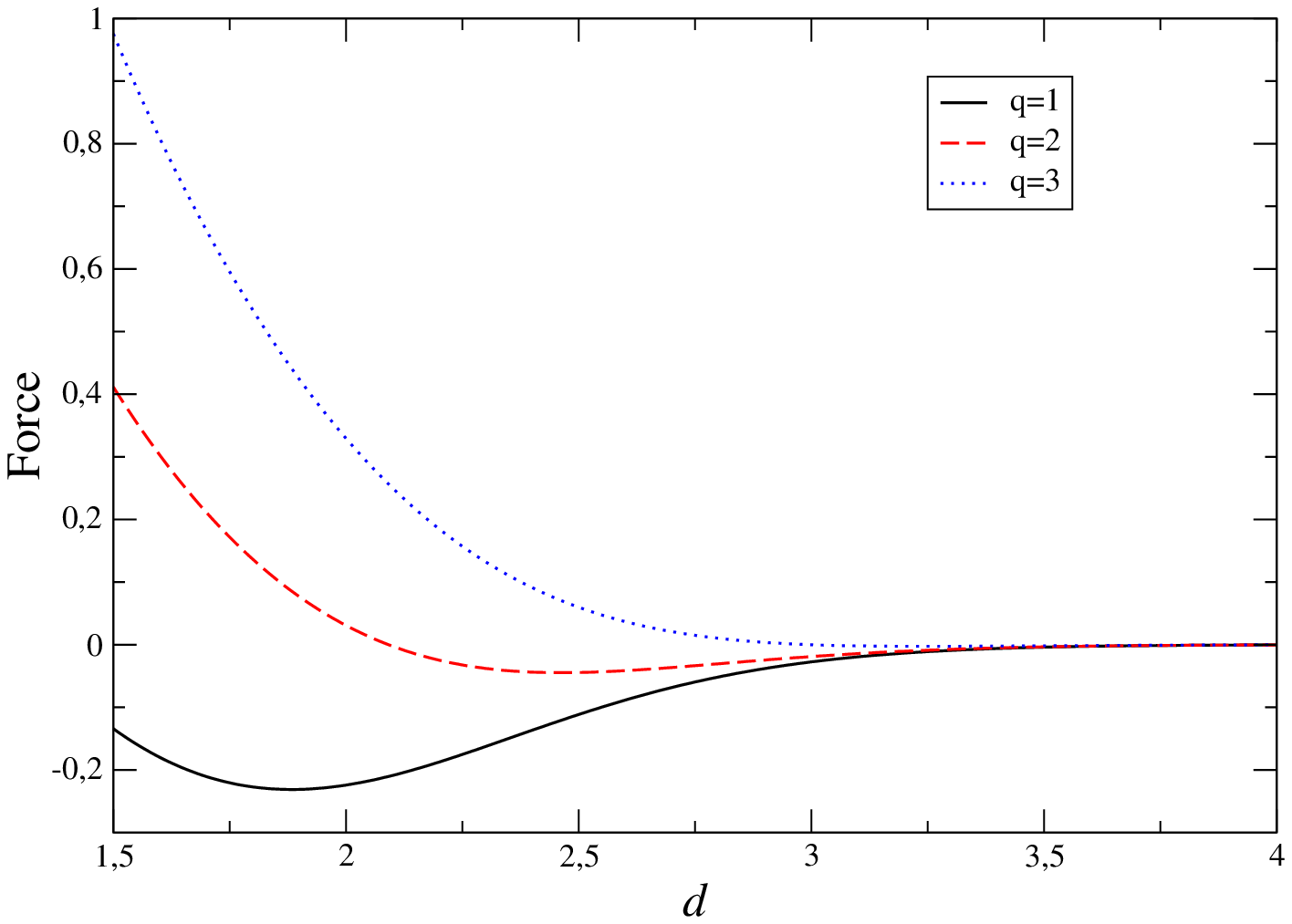}
\caption{The force between the guest charge $q e$ with the hard core of radius
$\sigma=1.5$ and the elementary guest charge $e$, immersed in the bulk of the
2D jellium at the free-fermion coupling $\Gamma=2$, versus the distance
between the guest charges $d\ge 1.5$.
The solid curve corresponds to $q=1$, the dashed curve to $q=2$ and
the dotted curve to $q=3$.}
\label{fig4}
\end{figure}

Numerical results for the force between the guest charge $q e$ with
the hard core of radius $\sigma=1.5$ and the elementary guest charge $e$,
obtained from the exact formula (\ref{forceg2}) by using {\it Mathematica},
versus the distance between the guest charges $d\ge 1.5$, are presented
in figure \ref{fig4}.
For the small value of the valence $q=1$ (solid curve), the force is
attractive for all distances $d$.
For $q=2$ (dashed curve), the guest particles repeal one another up to
the distance $d_t=2.09\ldots$ between them and then the mutual attraction
takes place up to an infinite distance, in accordance with the asymptotic
prediction (\ref{forceasym}).
The same scenario occurs for $q=3$ (dotted curve), the distance at which
the repulsion/attraction transition occurs is now $d_t=2.99\ldots$. 

\begin{figure}[tbp]
\centering
\includegraphics[clip,angle=-90,width=0.8\textwidth]{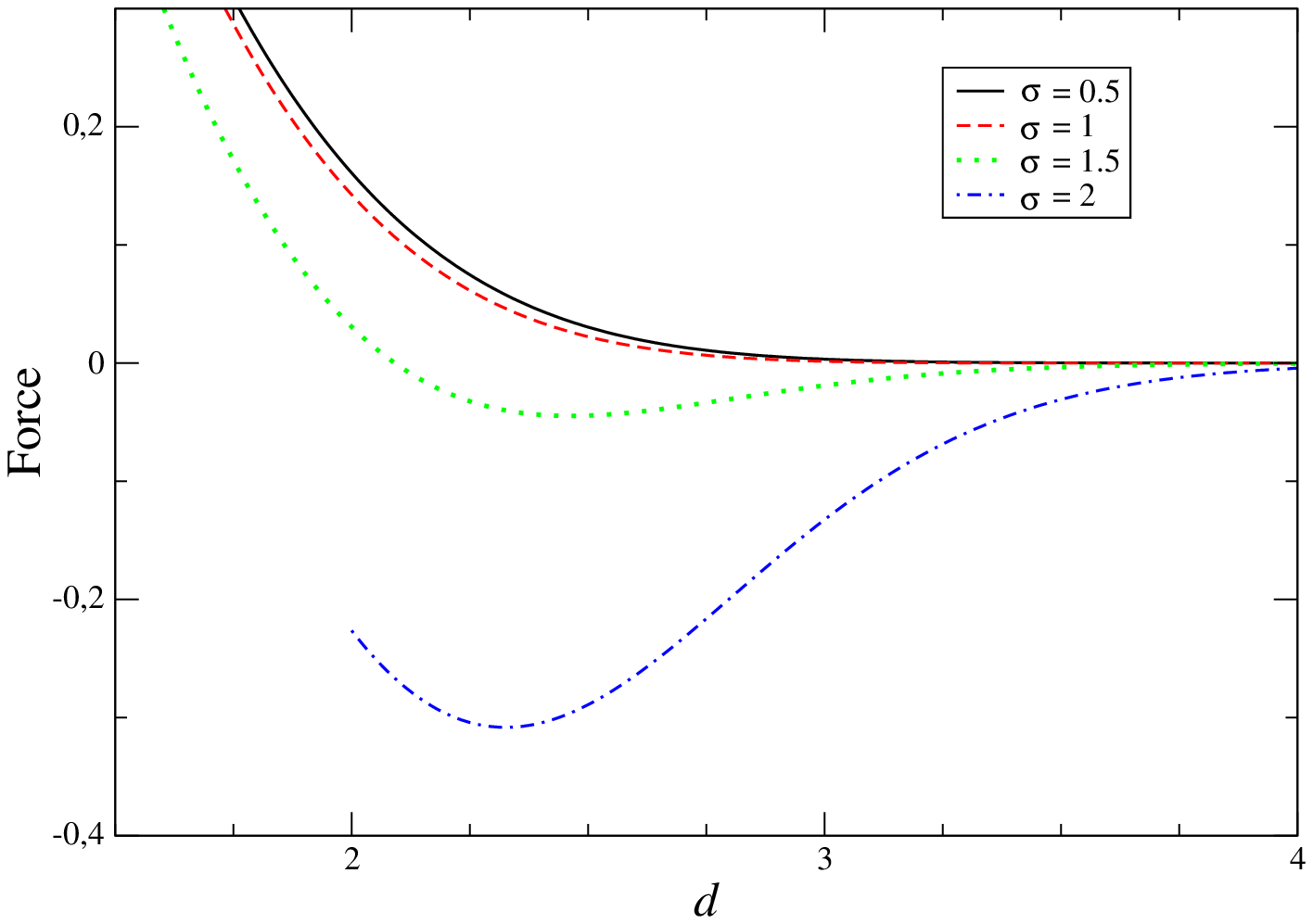}
\caption{The force between the guest charge $2 e$ with the hard core of radius
$\sigma$ and the elementary guest charge $e$, immersed in the bulk of the
2D jellium at the free-fermion coupling $\Gamma=2$, versus the distance
between the guest charges $d\ge\sigma$.
The solid curve corresponds to $\sigma=0.5$, the dashed curve to $\sigma=1$,
the dotted curve to $\sigma=1.5$ and the dotted-dashed curve to $\sigma=2$.}
\label{fig5}
\end{figure}

Numerical results for the force between the guest charge $2 e$ with
the hard core of radius $\sigma$ and the elementary guest charge $e$,
obtained from the exact formula (\ref{forceg2}) by using {\it Mathematica},
versus the distance between the guest charges $d\ge \sigma$, are presented
in figure \ref{fig5}.
In the case of small hard-core radiuses $\sigma=0.5$ (solid curve)
and $\sigma=1$ (dashed curve), for which the plots are very close to one
another, the force is repulsive for small distances $0<d<d_t$ and
attractive for $d\ge d_t$; in particular,
$d_t=4.80\ldots$ for $\sigma=0.5$ and $d_t=3.43\ldots$ for $\sigma=1$.
The same scenario holds for the intermediate $\sigma=1.5$ (dotted curve)
with $d_t=2.09\ldots$.
The force is attractive for all distances for the large hard-core
radius $\sigma=2$ (dash-dotted curve), i.e., $d_t=0$.

\renewcommand{\theequation}{7.\arabic{equation}}
\setcounter{equation}{0}

\section{Conclusion} \label{Sec7}
The motivation for the present work comes from the paper \cite{Ma01}
dealing with the 2D jellium of pointlike particles with the elementary charge
$e$ in thermal equilibrium at the inverse temperature $\beta$.
In that work, the effective interaction between two elementary guest charges
$e$ with the hard core of radius $\sigma$, immersed in the bulk of the jellium
at distance $d\ge \sigma$, was studied at the free-fermion coupling
$\Gamma\equiv\beta e^2 = 2$.
Because the fermion overlap matrix is non-diagonal in the presence of the
guest charges, the problem was treated perturbatively only.
In the leading order of the expansion in $\sigma$, the effective interaction
between likely charged guest charges is dominated by the repulsive Coulomb
interaction at small distances $d$ and by the charge-void attraction at
large distances $d$.
The crucial question is whether this phenomenon extends
to higher orders of the expansion in $\sigma$.

The only way how to answer this question is to solve exactly a specific model
in which at least one of the guest particles has the hard core, keeping in
this way the effect of the charge-void attraction.
We started our analysis by section \ref{Sec3} where thermodynamic quantities
in the presence of unit guest charges (with a non-diagonal overlap matrix)
are expressed in terms of multi-particle densities of the homogeneous
and translationally invariant jellium
(with a diagonal overlap matrix, see section \ref{Sec4}).
The crucial model we proposed in section \ref{Sec6} consists in fixing one guest
charge $q e$ ($q$ being any integer) with the hard core of radius $\sigma$
at the symmetry center of the neutralising background, preserving
in this way the diagonal form of the fermion overlap matrix.
In order to take advantage of the formalism developed in section \ref{Sec3},
the other guest particle must possess the elementary charge $e$ and
must be pointlike.
In the case of the pointlike ($\sigma=0$) guest charges $q e$ ($q=1,2,\ldots$)
and $e$ (section \ref{Sec61}), the effective force between them
(\ref{forceqe}) is always repulsive.
If the guest charge $q e$ ($q=0,\pm 1,\pm 2,\ldots$) has the hard core of
radius $\sigma>0$ (section \ref{Sec62}), the exact formula for
the effective force (\ref{forceg2}) leads to the attraction at asymptotically
large (finite) distances (\ref{forceasym});
we emphasize that all expansion orders of the hard-core radius $\sigma$
were taken into account in the derivation of this result.
This means that the phenomenon of dominancy of ion-void attraction at large
distances between the guest charges takes place also in our simplified model.
Although our model with hard core attached only to one guest charge
is less realistic than the one considered in \cite{Ma01}, it has advantage
of being exactly solvable, preserving at the same time the relevant
phenomenon of like-charge attraction.
The plots of the effective force as the function of the distance between
the guest charges are presented in figures \ref{fig4} and \ref{fig5}.

In the high-temperature $\Gamma\to 0$ limit, within the Debye-H\"uckel
theory the configuration of two guest charges $q_1 e$ and $q_2 e$
at distance $d$ induces in the jellium fluid the excess charge density
(\ref{chargedens}), see figure \ref{fig1}, whose asymptotic large-distance
limit (\ref{separation}) exhibits the separation form in the distance
and the corresponding angle.
This is no longer true for the free-fermion coupling $\Gamma=2$,
see figure \ref{fig2}.
The exact formula (\ref{cloud}), obtained as a by-product of the present
formalism, has a more complicated form which mixes the distance
and angle variables.

As to our potential plans for near future, it might be possible to obtain
exact results for the effective interaction between two guest charges
for higher coupling constants $\Gamma=4, 6,\ldots$.
It was mentioned in the Introduction that there exist algebraic techniques
for writing down the free energy of jellium systems, confined to a finite disc
or the surface of a cylinder, up to $N=11$ particles for $\Gamma=4$ and
$N=9$ particles for $\Gamma=6$ \cite{Tellez99,Tellez12,Samaj95,Samaj04,Samaj15}.
The necessary condition is that the fermion overlap matrix be diagonal
which might be attained by putting the two guest charges in appropriate
geometric configurations.
An analogous task has already been accomplished for 2D Coulomb systems
with counterions only possessing modulated surface charge densities
\cite{Samaj22}.

\ack
I am grateful to Emmanuel Trizac for pointing out my attention to
references \cite{Chapot04,Chapot05}. 
The support received from the project EXSES APVV-20-0150 and VEGA Grants
Nos. 2/0092/21 and 2/0089/24 is acknowledged.


\bigskip

\begin{thebibliography}{10}

\bibitem{Levin02} Levin Y 2002
Electrostatic correlations: from Plasma to Biology
{\it Rep. Prog. Phys.} {\bf 65} 1577--1632

\bibitem{Martin88} Martin Ph A 1988
Sum rules in charged fluids
{\it Rev. Mod. Phys.} {\bf 60} 1075--1127

\bibitem{Shklovskii99a} Shklovskii B I 1999
Wigner crystal model of counterion induced bundle formation of rodlike
polyelectrolytes  
{\it Phys. Rev. Lett.} {\bf 82} 3268--3271

\bibitem{Andelman06} Andelman D 2006
Introduction to electrostatics in soft and biological matter
In: Poon, W.C.K., Andelman, D. (eds.) {\it Soft Condensed Matter Physics
in Molecular and Cell Biology} vol. 6.
(Taylor \& Francis, New York)

\bibitem{Attard88} Attard P, Mitchell D J and Ninham B W 1988
Beyond Poisson-Boltzmann: Images and correlations in the electric double
layer. I. Counterions only
{\it J. Chem. Phys.} {\bf 88} 4987--4996

\bibitem{Netz00} Netz R R, Orland H 2000
Beyond Poisson-Boltzmann: Fluctuation effects and correlation functions
{\it Eur. Phys. J.} E {\bf 1} 203--214

\bibitem{Podgornik90} Podgornik R 1990
An analytic treatment of the first-order correction to the Poisson-Boltzmann
interaction free energy in the case of counter-ion only Coulomb fluid
{\it J. Phys. A: Math. Gen.} {\bf 23} 275--284

\bibitem{Verwey48} Verwey E J W and Overbeek J T G 1948
{\it Theory of the Stability of Lyophobic Colloids}
(Amsterdam: Elsevier)  
  
\bibitem{Chapot04} Chapot D, Bocquet L and Trizac E 2004
Interaction between charged anisotropic macromolecules:
Application to rod-like polyelectrolytes
{\it J. Chem. Phys.} {\bf 120} 3969--3982

\bibitem{Chapot05} Chapot D, Bocquet L and Trizac E 2005
Interaction between charged anisotropic macromolecules:
Application to rod-like polyelectrolytes
{\it J. Colloid Interface Sci.} {\bf 285} 609--618
  
\bibitem{Khan85} Khan A, J\"onsson B and Wennerstr\"om H 1985 
Phase equilibria in the mixed sodium and calcium di-2-ethylhexylsulfosuccinate 
aqueous system. An illustration of repulsive and attractive double-layer forces
{\it J. Phys. Chem.} {\bf 89} 5180--5184

\bibitem{Kjellander88} Kjellander R, Mar\v{c}elja S and Quirk J P 1988 
Attractive double-layer interactions between calcium clay particles
{\it J. Colloid Interface Sci.} {\bf 126} 194--211

\bibitem{Bloomfield91} Bloomfield V A 1991 
Condensation of DNA by multivalent cations: Considerations on mechanism
{\it Biopolymers} {\bf 31} 1471--1481

\bibitem{Kekicheff93} K\'ekicheff P, Mar\v{c}elja S, Senden T J and 
Shubin V E 1993 
Charge reversal seen in electrical double layer interaction of surfaces 
immersed in 2:1 calcium electrolyte 
{\it J. Chem. Phys.} {\bf 99} 6098--6113

\bibitem{Dubois98} Dubois M, Zemb T, Fuller N, Rand R P and 
Pargesian V A 1998
Equation of state of a charged bilayer system: Measure of the entropy of 
the lamellar–lamellar transition in DDABr 
{\it J. Chem. Phys.} {\bf 108} 7855--7869

\bibitem{Gulbrand84} Gulbrand L, J\"onsson B, Wennerstr\"om H and Linse P 1984 
Electrical double layer forces. A Monte Carlo study
{\it J. Chem. Phys.} {\bf 80} 2221-2228

\bibitem{Kjellander84} Kjellander R and Mar\v{c}elja S 1984 
Correlation and image charge effects in electric double-layers
{\it Chem. Phys. Lett.} {\bf 112} 49--53

\bibitem{Gronbech97} Gr{\o}nbech-Jensen N, Mashl R J, Bruinsma R F and
Gelbart W M 1997 
Counterion-Induced attraction between rigid polyelectrolytes
{\it Phys. Rev. Lett.} {\bf 78} 2477--2480 

\bibitem{Boroudjerdi05} Boroudjerdi H, Kim Y-W, Naji A, Netz R R, 
Schlagberger X and Serr A 2005
Statics and dynamics of strongly charged soft matter
{\it Phys. Rep.} {\bf 416} 129--199

\bibitem{Naji13} Naji A, Kandu\v{c} M, Forsman J and Podgornik R 2013
Perspective: Coulomb fluids -- Weak coupling, strong coupling, in between and 
beyond
{\it J. Chem. Phys.} {\bf 139} 150901

\bibitem{Moreira00} Moreira A G and Netz R R 2000 
Strong-coupling theory for counter-ion distributions
{\it Europhys. Lett.} {\bf 52} 705--711

\bibitem{Moreira01} Moreira A G and Netz R R 2001 
Binding of similarly charged plates with counterions only
{\it Phys. Rev. Lett.} {\bf 87} 078301

\bibitem{Netz01} Netz R R 2001
Electrostatics of counter-ions at and between planar charged walls: from
Poisson-Boltzmann to the strong-coupling theory
{\it Eur. Phys. J.} E {\bf 5} 557--574

\bibitem{Moreira02} Moreira A G and Netz R R 2002 
Simulations of counterions at charged plates
{\it Eur. Phys. J.} E {\bf 8} 33--58

\bibitem{Kanduc07} Kandu\v{c} M and Podgornik R 2007 
Electrostatic image effects for counterions between charged planar walls
{\it Eur. Phys. J.} E {\bf 23} 265--274

\bibitem{Shklovskii99b} Shklovskii B I 1999 
Screening of a macroion by multivalent ions: 
Correlation-induced inversion of charge
{\it Phys. Rev.} E {\bf 60} 5802--5811

\bibitem{Levin99} Levin Y, Arenzon J J and Stilck J F 1999 
The nature of attraction between like-charged rods
{\it Phys. Rev. Lett.} {\bf 83} 2680

\bibitem{Grosberg02} Grosberg A Y, Nguyen T T and Shklovskii B I 2002 
Colloquium: The physics of charge inversion in chemical and biological systems
{\it Rev. Mod. Phys.} {\bf 74} 329--345

\bibitem{Samaj11a} \v{S}amaj L and Trizac E 2011
Counterions at highly charged interfaces: From one plate to like-charge
attraction
{\it Phys. Rev. Lett.} {\bf 106} 078301

\bibitem{Samaj11b} \v{S}amaj L and Trizac E 2011
Wigner-crystal formulation of strong-coupling theory for counterions
near planar charged interfaces
{\it Phys. Rev.} E {\bf 24} 041401

\bibitem{Samaj16} \v{S}amaj L, dos Santos A P, Levin Y and Trizac E 2016 
Mean-field beyond mean-field: the single particle view for moderately 
to strongly coupled charged fluids
{\it Soft Matter} {\bf 12} 8768--8773

\bibitem{Palaia18} Palaia I, Trulsson M, \v{S}amaj L and Trizac E 2018
A correlation-hole approach to the electric double layer with counter-ions
only
{\it Mol. Phys.} {\bf 116} 3134--3146

\bibitem{Nordholm84} Nordholm S 1984
Simple analysis of the thermodynamic properties of the one-component plasma
{\it Chem. Phys. Lett.} {\bf 105} 302--307

\bibitem{Forsman04} Forsman J 2004 
A simple correlation-corrected Poisson-Boltzmann theory
{\it J. Phys. Chem.} B {\bf 108} 9236--9245

\bibitem{Baus80} Baus M and Hansen J P 1980 
Statistical mechanics of simple Coulomb systems
{\it Phys. Rep.} {\bf 59} 1--94 

\bibitem{Jancovici81} Jancovici B 1981
Exact results for the two-dimensional one-component plasma  
{\it Phys. Rev. Lett.} {\bf 46} 386--388.

\bibitem{Alastuey81} Alastuey A and Jancovici B 1981 
On the classical two-dimensional one-component Coulomb plasma
{\it J. Physique} {\bf 42} 1--12

\bibitem{Jancovici92} Jancovici B 1992
Inhomogeneous two-dimensional plasmas {\it Inhomogeneous Fluids}
ed D Henderson (New York: Dekker) pp 201--237

\bibitem{Forrester98} Forrester P J 1998 
Exact results for two-dimensional Coulomb systems
{\it Phys. Rep.} {\bf 301} 235--270 

\bibitem{Tellez99} T\'ellez G and Forrester P J 1999 
Exact finite-size study of the 2d-OCP at $\Gamma=4$ and $\Gamma=6$ 
{\it J. Stat. Phys.} {\bf 97} 489--521

\bibitem{Tellez12} T\'ellez G and Forrester P J 2012 
Expanded Vandermonde powers and sum rules for the two-dimensional 
one-component plasma
{\it J. Stat. Phys.} {\bf 147} 825--855

\bibitem{Samaj95} \v{S}amaj L and Percus J K 1995
A functional relation among the pair correlations of the two-dimensional
one-component plasma
{\it J. Stat. Phys.} {\bf 80} 811--824

\bibitem{Samaj04} \v{S}amaj L 2004
Is the two-dimensional one-component plasma exactly solvable?
{\it J. Stat. Phys.} {\bf 117} 131--158

\bibitem{Samaj15} \v{S}amaj L 2015
Counter-ions near a charged wall: Exact results for disc and planar geometries
{\it J. Stat. Phys.} {\bf 161} 227--249

\bibitem{Grimaldo15} Grimaldo J A M and T\'ellez G 2015 
Relations among two methods for computing the partition function of the 
two-dimensional one-component plasma
{\it J. Stat. Phys.} {\bf 160} 4--28

\bibitem{Ma01} Ma N, Girvin S M and Rajaraman R 2001
Effective attraction between like-charged colloids in a two-dimensional
plasma  
{\it Phys. Rev.} E {\bf 63} 021402

\bibitem{Gradshteyn} Gradshteyn I S and Ryzhik I M 2000 
{\it Table of Integrals, Series, and Products} 6th edn
(London: Academic Press)

\bibitem{Caillol81} Caillol J M 1981
Exact results for a two-dimensional one-component plasma on a sphere
{\it J. Phys. Lett. (France)} {\bf 42} 245--247  

\bibitem{Choquard83} Choquard Ph, Forrester P J and Smith E R 1983
The two-dimensional one-component plasma at $\Gamma=2$: the semiperiodic strip  
{\it J. Stat. Phys.} {\bf 33} 13--22

\bibitem{Alastuey84} Alastuey A and Lebowitz J L 1984
The two-dimensional one-component plasma in an inhomogeneous background:
exact results
{\it J. Phys. (France)} {\bf 45} 1859--1874

\bibitem{Cornu88} Cornu F and Jancovici B 1988
Two-dimensional Coulomb systems: a larger class of solvable models
{\it Europhys. Lett.} {\bf 5} 125--128

\bibitem{Berezin66} Berezin F A 1966
{\it The method of second quantization}  
(New York: Academic Press)

\bibitem{Samaj22} \v{S}amaj L 2022
Electric double layers with modulated surface charge density:
exact 2D results  
{\it J. Phys. A: Math. Theor.} {\bf 55} 275001

\end{thebibliography}
\end{document}